%% file: Psi_main.tex
\PassOptionsToPackage{draft=true}{hyperref}
\documentclass[fleqn,usenatbib]{mnras}

\usepackage{mathptmx}

\usepackage[T1]{fontenc}
\usepackage{ae,aecompl}
\usepackage{adjustbox}
\usepackage{array,booktabs}
\usepackage{multicol}        
\usepackage{multirow}        
\usepackage{float}
\usepackage{perpage}
\usepackage{epsfig,amstext,floatflt,alltt,fancyhdr,setspace}
\usepackage{graphicx}
\usepackage{lmodern} 
\usepackage{color}
\usepackage{ucs}
\usepackage{epstopdf}
\usepackage[utf8x]{inputenc}
\usepackage{url}
\usepackage{multirow}
\usepackage{mathrsfs}
\usepackage{amsmath} 
\usepackage{amssymb}
\usepackage{amsbsy}
\usepackage{appendix}
\usepackage{varioref}
\usepackage{times}

\usepackage{enumerate}

\usepackage[mathcal]{eucal}

\definecolor{purple}{RGB}{76, 0,153}

\def\ts{\thinspace}
\def \Eqt{Eq.\thinspace}
\def \Eqs{Eqs.\thinspace}
\def \sect{Section\thinspace}
\def \fig{Fig.\thinspace}
\def \figs{Figs.\thinspace}
\def \tab{Table\thinspace}
\def \App{Appendix\thinspace}

\def\d{{\rm d}}

\def\gg{\rm gg}
\def\gm{\rm gm}
\def\tb{\Bar{\theta}}

\def\DS8{\Delta\mathcal{S}_8}
\def\DSp{\Delta\mathcal{S}_8^{\perp}}
\def\Sp{\mathcal{S}_8^{\perp}}
\def\S8{\mathcal{S}_8}
\def\sigSp{\sigma_{\mathcal{S}_8^{\perp}}}
\def\sigS8{\sigma_{\mathcal{S}_8}}

\newcommand{\Om}{{\Omega_{\rm m}}}

\title[Scale-dependent galaxy bias and two-point statistics of the LSS]{Minimising the impact of scale-dependent galaxy bias on the joint cosmological analysis of large scale structures}

\author[Asgari, et al.]{Marika Asgari$^{1}$\thanks{E-mail: ma@roe.ac.uk},
Indiarose Friswell$^{1}$,
Mijin Yoon$^{2,3}$,
Catherine Heymans$^{1,2}$,\newauthor 
Andrej Dvornik$^{2}$,
Benjamin Joachimi$^{4}$,
Patrick Simon$^{5}$,
Joe Zuntz$^{1}$
\\
$^{1}$Institute for Astronomy, University of Edinburgh, Royal Observatory, Blackford Hill, Edinburgh, EH9 3HJ, UK\\
$^{2}$Ruhr-University Bochum, Astronomical Institute, German Centre for Cosmological Lensing, Universitätsstr. 150, 44801 Bochum, 	Germany\\
$^{3}$Department of Astronomy, Yonsei University, Yonsei-ro 50, Seoul, Korea \\
$^{4}$Department of Physics and Astronomy, University College London, Gower Street, London WC1E 6BT, UK\\
$^{5}$Argelander-Institut für Astronomie, Auf dem Hügel 71, 53121 Bonn, Germany
}

\date{Accepted XXX. Received YYY; in original form ZZZ}

\pubyear{2020}

\begin{document}

\label{firstpage}
\pagerange{\pageref{firstpage}--\pageref{lastpage}}
\maketitle

\begin{abstract}
We present a mitigation strategy to reduce the impact of non-linear
galaxy bias on the joint  `$3 \times 2 $pt' cosmological analysis of
weak lensing and galaxy surveys.  The $\Psi$-statistics that we adopt
are based on Complete Orthogonal Sets of E/B Integrals (COSEBIs).  As
such they are designed to minimise the contributions to the observable
from the smallest physical scales where models are highly uncertain.
We demonstrate that $\Psi$-statistics carry the same constraining
power as the standard two-point galaxy clustering and galaxy-galaxy
lensing statistics, but are significantly less sensitive to
scale-dependent galaxy bias.  Using two galaxy bias models, motivated
by halo-model fits to data and simulations, we quantify the error in a
standard $3 \times 2$pt analysis where constant galaxy bias is
assumed.  Even when adopting conservative angular scale cuts, that degrade the
overall cosmological parameter constraints,  we find of order $1 \sigma$ biases for
Stage III surveys on the cosmological parameter $S_8 =
\sigma_8(\Om/0.3)^{\alpha}$.  This arises from a leakage of the
smallest physical scales to all angular scales in the standard
two-point correlation functions.  In contrast, when analysing
$\Psi$-statistics under the same approximation of constant galaxy
bias, we show that the bias on the recovered value for $S_8$ can be
  decreased by a factor of $\sim 2$, with less conservative scale cuts. Given the
challenges in determining accurate galaxy bias models in the highly
non-linear regime, we argue that $3 \times 2$pt analyses should move
towards new statistics that are less sensitive to the smallest
physical scales.

\end{abstract}




\section{Introduction}
\input{Introduction.tex}

\section{Galaxy bias}
\input{Bias_Model.tex}

\section{Two point statistics}
\input{Methods.tex}

\section{Cosmological models and survey setups}
\input{Surveys.tex}

\section{The impact of scale-dependent galaxy bias on cosmological analysis}
\input{sensitivity.tex}

\vspace{0.5cm}

\section{Summary and conclusions}
\input{Conclusions.tex}

\section*{Acknowledgements}
{\small
We used chainconsumer \citep{Hinton2016} for making \figs\ref{fig:KiDS-BOSS-like_psi} and \ref{fig:DES-Y1-5000_psi}.
We acknowledge support from the European Research Council under grant
agreement No.~647112 (CH, MA), and No.~770935 (AD). CH and MY acknowledge support from the Max Planck Society and the Alexander von Humboldt Foundation in the framework of the Max Planck-Humboldt Research Award endowed by the Federal Ministry of Education and Research. MY acknowledges support from the National Research Foundation (NRF) of Korea grant funded by the Korea government (MSIT) under no.2019R1C1C1010942.
}

\section*{Data Availability}
{\small
No new data were generated in support of this research.
}

\newpage
\bibliographystyle{mnras}
\bibliography{Psi_main}

\onecolumn
\newpage
\appendix

\section{Covariance of $\Psi$-statistics}
\input{covariance.tex}

\section{Additional figures}
\input{appendix_fig.tex}

\bsp	
\label{lastpage}

\end{document}

%% file: Introduction.tex
\label{sec:Introduction}

Combined analysis of weak gravitational lensing and galaxy surveys has recently become a standard approach for analysing cosmological data. 
This approach uses three sets of two-point statistics ($3\times2$pt for short), characterising cosmic shear, galaxy clustering and the cross correlation between galaxy positions and background shear, known as galaxy-galaxy lensing (GGL). 
This combination of probes, allows for improved constraints on cosmological parameters through degeneracy breaking between both cosmological and nuisance parameters \citep{DES, Joudaki_KiDS_2dFLenS, vanuitert/etal:2018}. 

%

Galaxies are biased tracers of the underlying matter distribution and any cosmological probe that relies on their positions has to take this bias into account. On very large physical scales the galaxy bias can be characterised by a constant, however, on smaller scales this relationship breaks and the galaxy distribution deviates from the matter distribution in a scale-dependent manner \citep{Bias_Review}. On quasi-linear scales, perturbation theory can be used to model this scale-dependence \citep[see for example][]{chan/etal:2012}, which is one method used to analyse the galaxy clustering signal of the Baryon Oscillation Spectroscopic Survey (BOSS) data \citep{gilmarin/etal:2016,beutler/etal:2017,grieb/etal:2017,sanchez/etal:2017,damico/etal:2019,ivanov/etal:2019,troster/etal:2020}. On smaller scales either a halo model approach or simulation results can be used to model galaxy bias \citep[for example][]{cacciato/etal:2012,springle/etal:2018}. 

The first series of $3\times2$pt analyses \citep{DES,vanuitert/etal:2018,Joudaki_KiDS_2dFLenS} applied scale cuts to their data and adopted a constant effective galaxy bias model. Depending on the chosen two-point statistics, however, the sensitivity to smaller physical scales, and hence the contribution from the scale-dependent bias, varies. \cite{DES} analysed the first year of data from the Dark Energy Survey \citep[DES,][]{DES_DataRelease1} using real space correlation functions. On the other hand \citet{vanuitert/etal:2018} analysed the combination of Galaxy And Mass Assembly \citep[GAMA,][]{GAMA} and the first 450 deg$^2$ of the Kilo Degree Survey \citep[KiDS,][]{Kuijken15,dejong/etal:2017} with angular power spectra. \cite{Joudaki_KiDS_2dFLenS} adopted the real space correlation function, $\gamma_{\rm t}$, for their GGL signal and redshift-space multipole power spectra for their clustering signal to analyse KiDS-450 with BOSS and the 2-degree Field Lensing Survey \citep[2dFLenS,][]{blake/etal:2016}.

In this paper we explore the sufficiency of scale cuts for the recovery of unbiased cosmological parameters, in analyses where the scale-dependence of galaxy bias is ignored.  We quantify the impact of scale-dependent galaxy bias on a $3\times2$pt analysis, using a Fisher formalism. In addition, we advocate the use of a different set of statistics, `$\Psi$-statistics', based on the Complete Orthogonal Sets of E/B-Integrals \citep[COSEBIs,][]{SEK10}. COSEBIs are statistics designed for cosmic shear analysis which are able to minimise the effect of small physical scales, while staying clear of the measurement challenges of Fourier space statistics \citep{asgari/etal:2019,asgari/etal:2020}.

$\Psi$-statistics were first proposed by \citet{Buddendiek} as an approach that is able to limit the angular scales used in the measurement. They applied this method to the combination of the Red Cluster Sequence Lensing Survey \citep[RCSLenS,][]{hildebrandt/etal:2016} and BOSS galaxies. They fixed the cosmological parameters and constrained two galaxy bias parameters assuming a constant bias model; the bias factor $b$, characterising the galaxy autocorrelation and $r$, the galaxy-matter cross-correlation coefficient. Here we consider three galaxy bias models: a scale-independent model characterised by a single scaling parameter, $b$, as well as two well-motivated scale-dependent models from \cite{Andrej_bias_model} and \cite{Simon_bias_model}. We compare the response and sensitivity of correlation functions, angular power spectra and $\Psi$-statistics to these bias models, using their signal-to-noise ratios. We consider two surveys, one corresponding to a DES year 1 survey scaled to have the same area as the final DES data release, and the other to the combination of BOSS and 1000 deg$^2$ of KiDS data.
For these surveys, we estimate the level of systematic errors on cosmological parameters introduced by neglecting the scale-dependence of galaxy bias. To find these errors, we produce mock data from the scale-dependent models, but analyse them with a constant bias model. We quantify the results for the parameter $S_8=\sigma_8(\Om/0.3)^{\alpha}$ and the combination of galaxy clustering and GGL.

In \sect\ref{sec:Bias_Model} we introduce the galaxy bias models that we explore in our analysis. We then describe the three sets of two-point statistics in \sect\ref{sec:Methods}. The cosmological model and the survey setups are detailed in \sect\ref{sec:surveys_cosmo}. Our results are shown in \sect\ref{sec:sensitivity_Fisher}. Finally, we  conclude in \sect\ref{sec:Conclusions}. The covariance matrix of $\Psi$-statistics is calculated in \App\ref{app:covariance}.

%% file: Bias_Model.tex
\label{sec:Bias_Model}

\begin{figure}
\begin{center}
     \begin{tabular}{c}
  \includegraphics[width=0.95\linewidth]{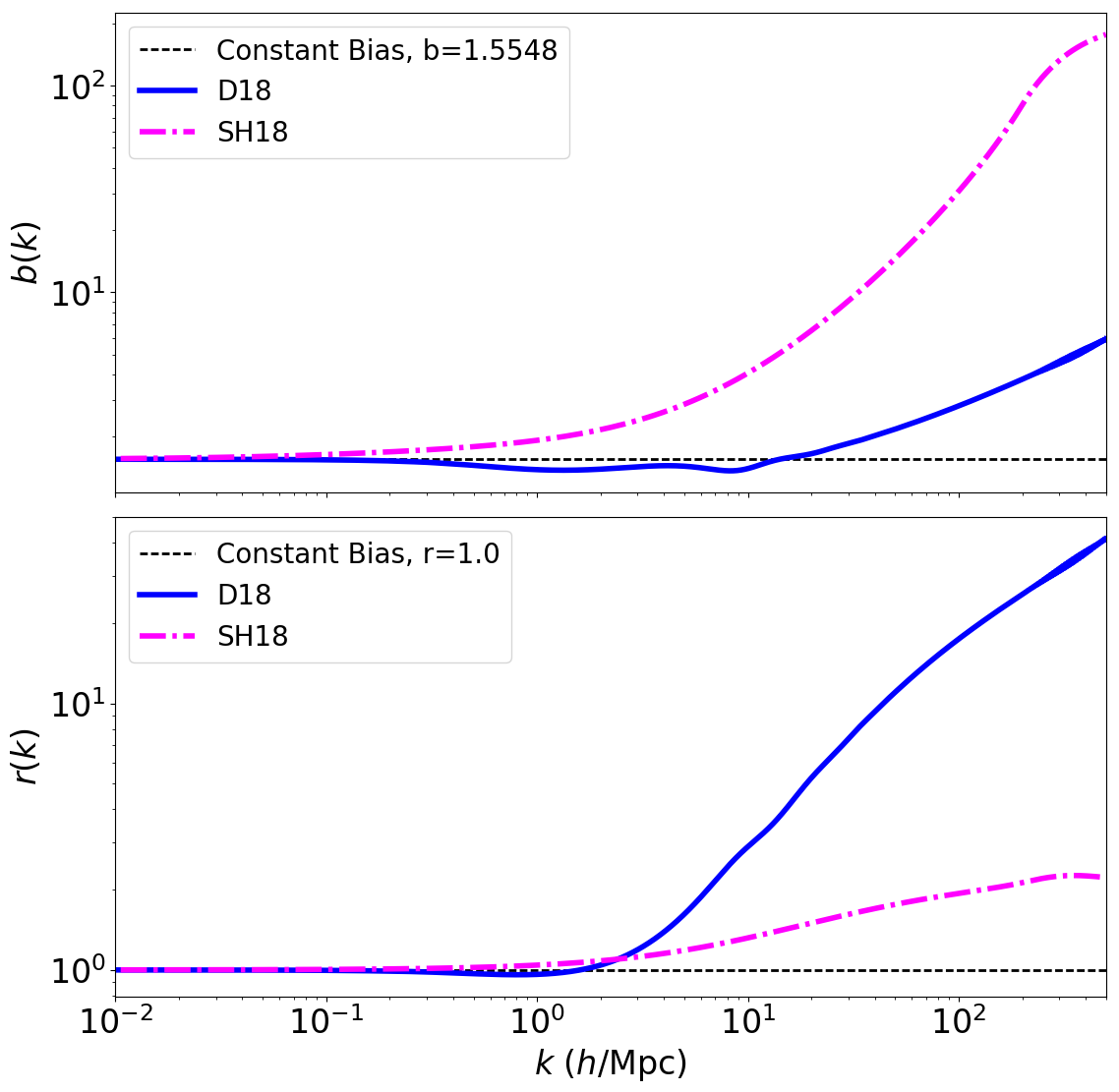}
  \end{tabular}
  \end{center}
\caption{Galaxy bias models: constant (dashed black), D18 \citep{Andrej_bias_model} (solid blue) and SH18 \citep{Simon_bias_model} (dot-dashed magenta). The bias function, $b(k)$, is shown on the top while $r(k)$ is shown on the bottom. All models have been scaled to the same constant bias at large scales, $r=1$ and $b=1.55$. We note that the models can only be trusted to $k\sim10\;h\,$Mpc$^{-1}$ and therefore, the value at higher $k$ are a result of extrapolations. They do, however, provide a useful test bed for the impact of uncertain galaxy bias at high $k$ on $3\times2$pt analysis. }
\label{fig:bias_models}
\end{figure}

Galaxy bias characterises the statistical relation between the distributions of galaxies and matter, which at the two-point level can be described by $b(k,z)$ and $r(k,z)$ as function of scale, $k$, and redshift, $z$. 
The bias function $b(k,z)$, expresses the fluctuation in the variance of the galaxy number density relative to the variance in the matter density \citep{tegmark/etal:1999}, 
\begin{equation}
b^{2}(k,z) = \frac{P_{\rm gg}(k,z)}{P_{\rm mm}(k,z)}\;,
\end{equation}
where $P_{\rm gg}(k,z)$ and $P_{\rm mm}(k,z)$  are the power spectra of galaxies and matter, respectively. The bias function $r(k,z)$, is a measure for the correlation between the galaxy and matter density \citep{dekel_lahav:1999},
\begin{equation}
r(k,z) = \frac{P_{\rm gm}(k,z)}{\sqrt{P_{\rm gg}(k,z)P_{\rm mm}(k,z)}}\;,
\end{equation}
where $P_{\rm gm}(k,z)$ is the cross-power spectrum between matter and galaxies.
The scale dependence of the galaxy bias is weak on large scales, $k\ll1\,h\,{\rm Mpc}^{-1}$ \citep[see for example][]{cresswell_percival:2009}. We, however, expect these functions to vary strongly with $k$ on non-linear scales, as predicted by perturbation theory, halo models and simulations \citep[see for example][]{Bias_Review,weinberg/etal:2004}. In addition, galaxy bias is specific to a galaxy population. For instance, luminous galaxies are more biased than faint galaxies.

In this work, we study the effect of the scale dependence of galaxy bias on cosmological analyses. We therefore assume no redshift evolution and thus skip the z-argument in the bias functions. This is because here we are only interested in how statistics with different scale dependence respond to galaxy bias. Therefore, we choose two galaxy populations with distinct models of $b(k)$ and $r(k)$ to represent the diversity of scale-dependent bias found in observations or simulations. We normalise these models to $r=1$ and $b=1.55$ at $k<0.01\,h\,{\rm Mpc}^{-1}$ to have the same deterministic bias on large scales and at the same time different variations with scale. The two models are as follows:

\begin{enumerate}
\item \textbf{SH18} - The first model is based on templates fitted by \cite{Simon_bias_model} to red galaxies at redshift of around $0.5$ which are similar to the BOSS high-redshift sample \citep{Reid}. Specifically, we use the polynomial functions,
\begin{equation}
b(k)=\frac{b_{0} + b_{1}k + b_{2}k^{2}}{1 +  b_{3}k^{3} + b_{4}k^{4}}\;, \ \ \ \ \ \ \ \ r(k)=\frac{1 + r_{1}k + r_{2}k^{2}}{1 +  r_{3}k + r_{4}k^{2} + r_{5}k^{3}}\;,
\label{eq:b_polynomial}
\end{equation}
where the values to $b_i$ and $r_i$ can be found using the numerical simulations. The galaxy bias of this sample is based on the semi-analytic model by \citet{henriques/etal:2015}. For the red BOSS-like sample the fit values are $b_0 =1.5548$, $b_1 = 11.3538$, $b_2 = 1.4771$, $b_3 = 6.5046$, $b_4 = -0.0135$, $r_1 =  0.0807$, $r_2 = -0.0002$, $r_3 = 0.0365$, $r_4 = -0.0001$ and $r_5 =  0.0$. The fit to the polynomial breaks at $k=200\,h\,{\rm Mpc}^{-1}$, therefore we smooth the function to a constant value close to the value of each bias function at $k=200 \;h/$Mpc for higher $k$.

\item \textbf{D18} - \cite{Andrej_bias_model} used a halo model approach based on \cite{cacciato/etal:2012} to model the scale dependence of galaxy bias and fit this model to the GAMA survey \citep[see also][who used a similar method on the Sloan Digital Sky Survey data]{zehavi/etal:2011}. They used the first 450 square degrees of the KiDS data as their sources to measure the GGL signal. Here we use the results from their highest of the three galaxy mass bins corresponding to the stellar mass range of 10.9-12.0 $\log (M_*/[{\rm M}_\odot/h^2])$. 
\end{enumerate} 

As an additional third model, we adopt a scale-independent bias that converges with  both SH18 and D18 on large scales. We explore the impact of a scale-dependent bias in our analysis by comparing the results to that of a constant bias. \fig\ref{fig:bias_models} shows the three galaxy bias models, constant (black dashed) and the two scale-dependent bias models, D18 (solid blue) and SH18 (dot-dashed magenta), in terms of Fourier modes, $k$. As expected they are all constant at small $k$-scales, but diverge as $k$ increases. For $b(k)$ we see that the SH18 model becomes scale-dependent at smaller $k$-scales compared to the D18 model and increases more rapidly with $k$. The correlation factor $r(k)$, on the other hand, departs from the constant value at smaller $k$ values for D18 than for SH18, while both are close to $r=1$ up to $k\sim1\,h\,{\rm Mpc}^{-1}$. The increase $r>1$ beyond this $k$ is probably related to the galaxies located at the centre of halos (central galaxies), or a non-Poisson variance of galaxy numbers inside matter halos \citep{guzik_seljak:2001}.

The values provided by these galaxy bias models at high $k$ are uncertain\footnote{It is also uncertain at which high $k$ the models become less reliable. However, we estimate that this happens at $k\sim 10$.}. These scales represent an extrapolation into a regime where neither the simulations nor the data directly constrain the model. However, they cannot be considered completely unreasonable as the halo model, used for semi-analytic galaxy models or in the analysis of D18, provides an excellent description of the galaxy population on scales where it can be directly tested.  These models therefore provide a reasonable test bed for the leakage of galaxy bias variations at high-$k$ into two-point statistics, especially since the behaviour of D18 and SH18 are very different at high $k$-scales.

%% file: Methods.tex
\label{sec:Methods}

In the following sections we introduce three sets of two point statistics for measuring the galaxy clustering and GGL signals. They are angular power spectra: $C^{\gg}(\ell)$ and $C^{\gm}(\ell)$, real-space correlation functions: $\omega(\theta)$ and $\gamma_{\rm t}(\theta)$, and the $\Psi$-statistics: $\Psi^{\gg}_{n}$ and $\Psi^{\gm}_{n}$. Here we show how each statistic is related to the underlying matter power spectrum. In \App\ref{app:covariance} we calculate the covariance of the $\Psi$-statistics.

\subsection{Angular power spectra: $C^{\gg}(\ell)$ and $C^{\gm}(\ell)$}
\label{sec:angular_power_spectra_section}

Theoretical models of cosmology generally provide us with the matter power spectrum, $P_{\rm mm}(k,\chi)$, which contains all the two-point statistical information about the matter distribution as a function of both scale, $k$, and co-moving distance, $\chi$, \citep{Kaiser_1998}. We can project this information into two-dimensional angular power spectra by projecting the matter distribution onto a two dimensional surface, by integrating over the $\chi$ dependence of $P_{\rm mm}(k,\chi)$.  We can measure the angular power spectra of both the galaxy-galaxy auto-correlation, $C^{\gg}(\ell)$ and galaxy-matter cross-correlation (GGL), $C^{\gm}(\ell)$, from the data. Galaxies are, however, biased tracers of the matter distribution and therefore to connect these power spectra to $P_{\rm mm}(k,\chi)$, we need to include the galaxy bias functions, $b(k,\chi)$ and $r(k,\chi)$, now written in terms of the co-moving distance, $\chi$ instead of redshift. We can then connect the angular power spectra to the three dimensional matter power spectrum using an extended Limber approximation \citep{LoverdeAfshordi08, kilbinger/etal:2017},
\begin{align}
\label{eq:P_gg}
        C^{\gg}(\ell)\ =\int_{0}^{\chi_{\rm h}} &\d\chi\ \frac{p_{\rm f}(\chi)^{2}}{f_{\rm{K}}(\chi)^{2}}\nonumber \\
        & \times\ b^{2}\left(k=\frac{\ell+0.5}{f_{\rm{K}}(\chi)};\chi\right)\ P_{\rm mm}\left(k=\frac{\ell+0.5}{f_{\rm{K}}(\chi)};\chi \right)\;, 
\end{align}
and
\begin{align}
& C^{\gm}(\ell)=\frac{3 \Omega_{\rm m} H^{2}_{0} } { 2c^{2} } \int_{0}^{\chi_{\rm h}} \d\chi\ \frac{\ p_{\rm f}(\chi)g(\chi)}{a(\chi) f_{\rm{K}}(\chi)}\nonumber \\
& \times b\left(k=\frac{\ell+0.5}{f_{\rm{K}}(\chi)};\chi\right) r\left(k=\frac{\ell+0.5}{f_{\rm{K}}(\chi)};\chi\right)\  P_{\rm mm}\left(k=\frac{\ell+0.5}{f_{\rm{K}}(\chi)};\chi \right)\;, 
\label{eq:P_gm}
\end{align}
where the integrals are evaluated from the co-moving distance $\chi=0$ to the co-moving distance to the horizon, $\chi_{\rm h}$. Here $p_{\rm f}(z)$ is the probability density of the foreground (lens) galaxies, $c$ is the speed of light, $f_{\rm{K}}(\chi)$ is the co-moving angular diameter distance, $a(\chi)$ is the scale factor, $H_{0}$ is the value of the Hubble constant today and $\Omega_{\rm m}$ is the matter density parameter.
The gravitational  lensing weight, $g(\chi)$, is given by
\begin{equation}
g(\chi)=\int_{\chi}^{\chi_{\rm h}} \d\chi' p_{\rm b}(\chi') \frac{ f_{\rm{K}}(\chi'-\chi)}{f_{\rm{K}}(\chi')}\;,
\label{eq:WFiltereq}
\end{equation}
where $p_{\rm b}(z)$ is the probability density of the background (source) galaxies.

\subsection{Real-space correlation functions: $\omega(\theta)$ and $\gamma_{\rm t}(\theta)$}
\label{sec:real_space_section}

The real space counterparts to $C^{\gg}(\ell)$ and $C^{\gm}(\ell)$ are the two point correlation functions, usually denoted as $\omega(\theta)$  and $\gamma_{\rm t}(\theta)$, respectively. These functions can be measured from the catalogues directly, and their predicted values can be written in terms of angular power spectra,
\begin{equation}
    \omega(\theta) = \int_{0}^{\infty} \frac{\ell\ \d\ell}{2\pi}\ {\rm J}_{0}(\ell\theta)\ C^{\gg}(\ell)\;,
    \label{eq:clustering}
\end{equation}
and
\begin{equation}
    \gamma_{\rm t}(\theta) = \int_{0}^{\infty} \frac{\ell\ \d\ell}{2\pi}\ {\rm J}_{2}(\ell\theta)\ C^{\gm}(\ell)\;,
    \label{eq:ggl}
\end{equation}
where ${\rm J}_{x}$ is the $x^{\rm th}$ order Bessel function of the first kind \citep[]{hu_jain:2004}.

%

%


\subsection{$\Psi$-statistics: $\Psi^{\gg}_{n}$ and $\Psi^{\gm}_{n}$}
\label{sec:psi_space_section}

\begin{figure*}
\centering
\begin{minipage}{.5\textwidth}
  \centering
  \includegraphics[width=\linewidth]{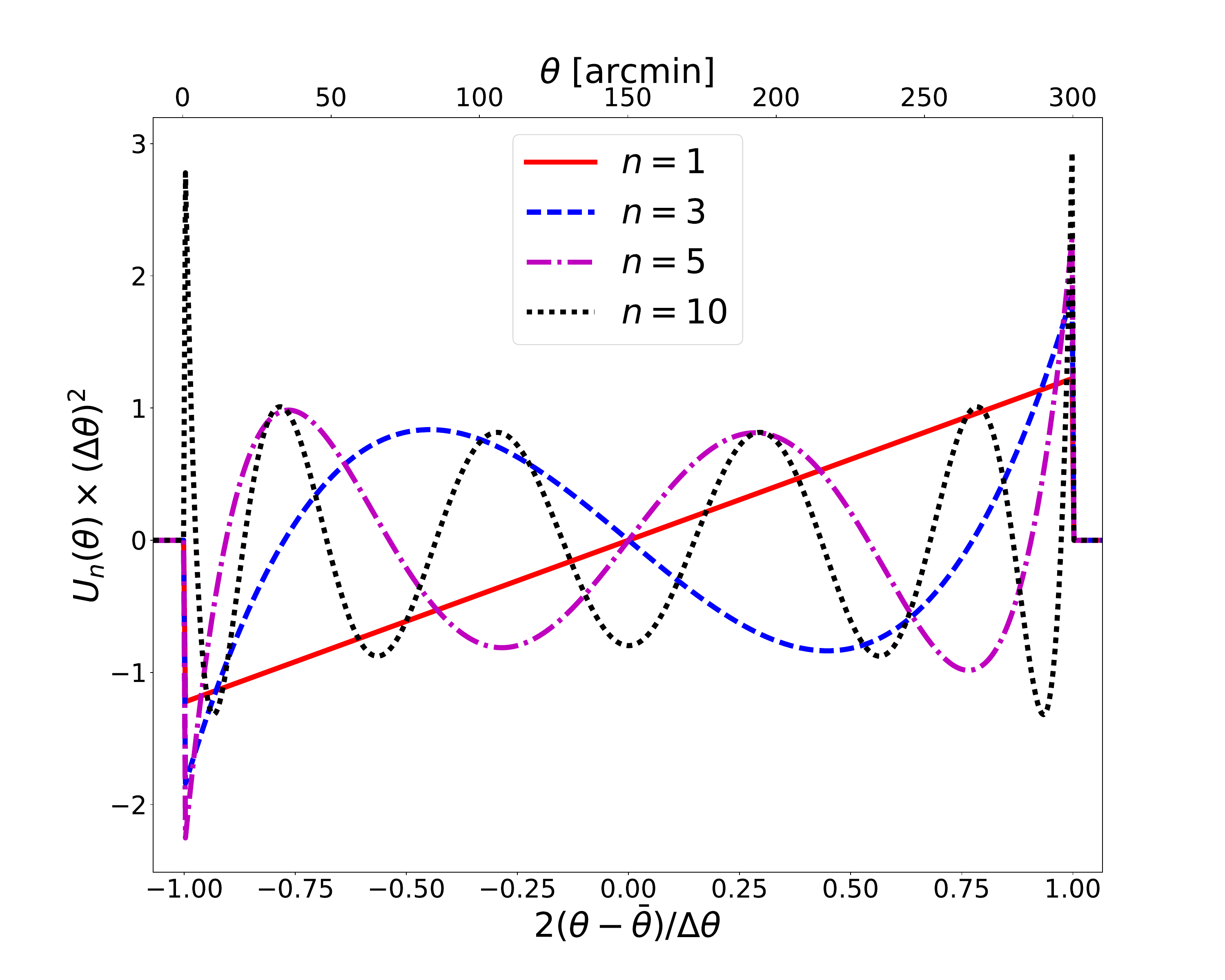}
\end{minipage}%
\begin{minipage}{.5\textwidth}
  \centering
  \includegraphics[width=\linewidth]{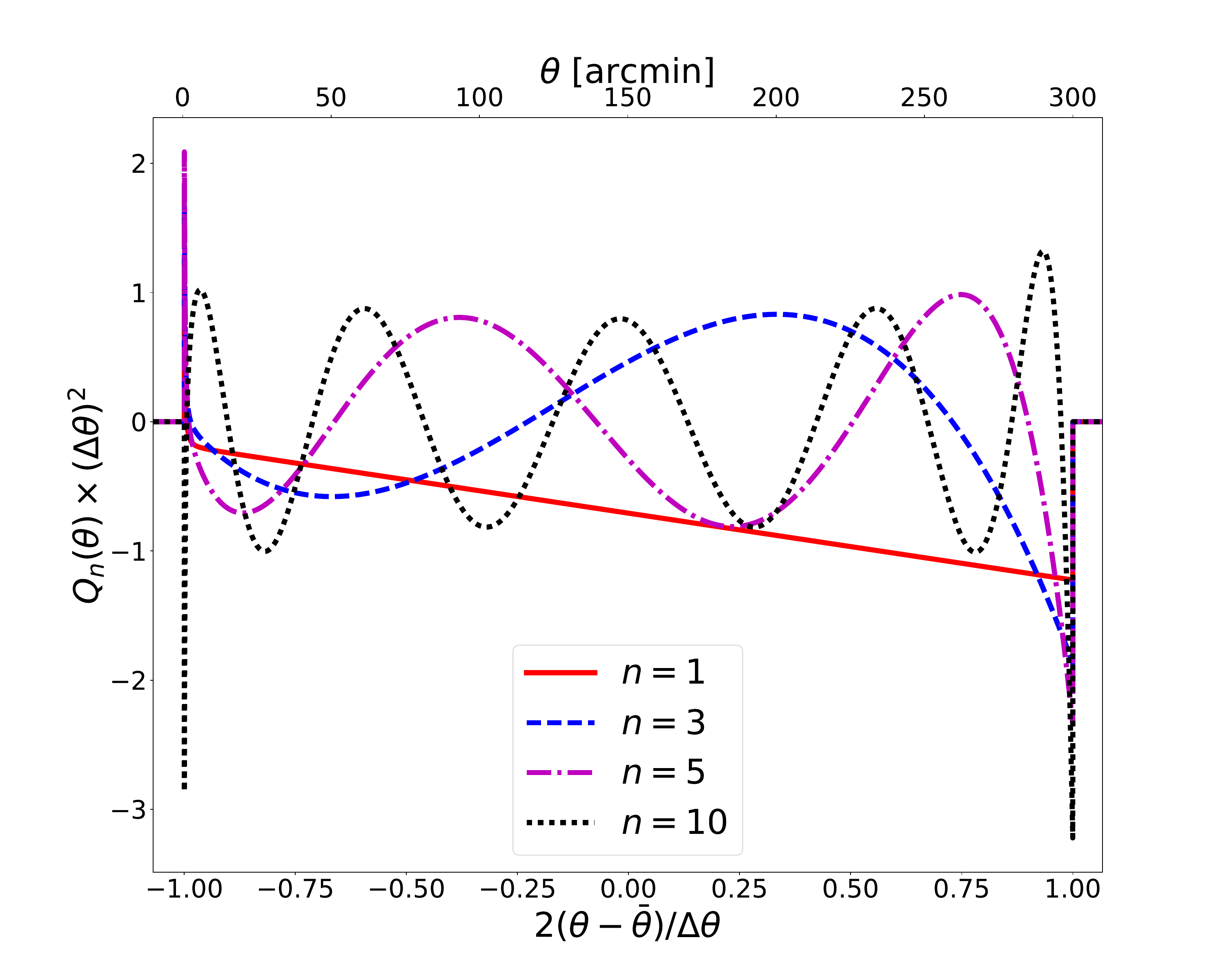}
\end{minipage}
\caption{Filter functions $U_{n}(\theta)$ (left, \Eqs\ref{eq:U1} and \ref{eq:Un}) and $Q_{n}(\theta)$ (right, \Eqs\ref{eq:Q1} and \ref{eq:Qn}). They are shown as a function of  $2(\theta-\bar{\theta})/\Delta\theta$, mapped on to the interval $[-1,1]$. The upper horizontal axis shows the mapping to $\theta$ for $\theta_{\rm min}=0.5'$ and  $\theta_{\rm max}=300'$. }
\label{fig:UQfilters}
\end{figure*}

Although the correlation functions can be measured more directly from the data, they mix information from high and low $\ell$ values, which complicates their modelling. 
Several alternative statistics have been proposed to the standard $\gamma_{\rm t}(\theta)$ and $\omega(\theta)$ observables. \cite{Baldauf}, for example, developed a set of statistics, $\hat{\Upsilon}(\theta,\theta_{\rm min})$, which aimed to dampen contributions from small scales for the GGL signal, given by
\begin{equation}
    \hat{\Upsilon}(\theta,\theta_{\rm min}) = \gamma_{\rm t}(\theta) - \left(\frac{\theta_{\rm min}}{\theta}\right)^{2}\gamma_{\rm t}(\theta_{\rm min})\;,
    \label{eq:OldUpsilon}
\end{equation}
where scales below $\theta_{\rm min}$ do not contribute to the measured signal. Initially, this approach was thought to solve many problems as it suppressed the information from small scales, below $\theta_{\rm min}$, which are not well understood. There are, however, a number of implementation challenges for this statistic. As $\gamma_{\rm t}(\theta_{\rm min})$ feeds into all data points, systematic measurement errors at $\theta_{\rm min}$  propagate to all scales. In addition, $\Upsilon$-statistics are only able to remove  scales below $\theta_{\rm}$ if the measured $\gamma_{\rm t}(\theta_{\rm min})$ is an unbiased estimate. In practice the data is binned in $\theta$ and therefore residual biases will be present in the estimate of $\gamma_{\rm t}(\theta_{\rm min})$ which will also affect $\Upsilon$ on all scales.

These issues were addressed in \cite{Buddendiek} who proposed a new set of statistics which they also called $\Upsilon^{\gg,\gm}_{n}$. Here we update their notation to $\Psi^{\gg,\gm}_{n}$ to avoid confusion with the \cite{Baldauf} statistics. \cite{Buddendiek} realised that the definition in \Eqt\eqref{eq:OldUpsilon} is a special case of aperture mass statistics \citep{schneider:1996} and therefore can be written as a weighted integral on $\gamma_{\rm t}$. They used the aperture mass formalism to generalise and improve the \cite{Baldauf} statistic. This resulted in $\Psi^{\gg,\gm}_{n}$ which are discrete functions and can be written with respect to the real space correlation functions,
\begin{equation}     
\Psi^{\gg}_{n} = \int_{\theta_{\rm min}}^{\theta_{\rm max}} \d\theta'\ \theta' \ U_{n}(\theta')\ \omega(\theta')\;,
\label{eq:psi_gg_data}
\end{equation}
and
\begin{equation}
\Psi^{\gm}_{n} = \int_{\theta_{\rm min}}^{\theta_{\rm max}} \d\theta'\ \theta' \ Q_{n}(\theta')\ \gamma_{\rm t}(\theta')\;,
\label{eq:psi_gm_data}
\end{equation}
where $U_{n}(\theta)$ and $Q_{n}(\theta)$ are filter functions defined on a finite angular range of $\theta\in [\theta_{\rm min},\theta_{\rm max}]$. The $U_{n}(\theta)$ functions are compensated, 
\begin{equation}
\int_{\theta_{\rm min}}^{\theta_{\rm max}} \d\theta'\ \theta'\ U_{n}(\theta')\ =\ 0\;,
\label{eq:compensated_condition}
\end{equation}
and are chosen to be orthogonal
\begin{equation}
\int_{\theta_{\rm min}}^{\theta_{\rm max}} \d\theta'\ U_{n}(\theta')\ U_{m}(\theta')\ =\ 0\ \ \ \rm for\  m\neq n\;.
\label{eq:orthogonal_condition}
\end{equation}
The $U_{n}(\theta)$ functions therefore form a complete set of filter functions which means that they contain all the information in their range of support, except a constant amplitude which is nulled as a result of \Eqt\eqref{eq:compensated_condition}. For a gravitational lensing signal this condition removes the ambiguity due to mass-sheet degeneracy. The $Q_{n}(\theta)$ functions can be calculated for each $U_{n}(\theta)$ using
\begin{equation}
Q_{n}(\theta) = \frac{2}{\theta^2} \int_{0}^{\theta}\d\theta'  \theta' U_{n}(\theta')   - U_{n}(\theta)\;.
\label{eq:Q_Filter_Function_integral}
\end{equation}
There are infinite families of $U_{n}(\theta)$ and $Q_{n}(\theta)$ functions that satisfy the conditions in Eqs.\ts\eqref{eq:compensated_condition}, \eqref{eq:orthogonal_condition} and \eqref{eq:Q_Filter_Function_integral}. \cite{Buddendiek} proposed using the Legendre polynomials,  $P_{n}$, with some modifications to form the $U_n(\theta)$ functions. The first mode, $n=1$, is defined as
\begin{equation}
U_1(\theta)=\frac{1}{(\Delta\theta)^3}\frac{12\tb(\theta-\tb)}{\sqrt{(\Delta\theta)^2+24\tb^2}}
\label{eq:U1}
\end{equation}
and the higher modes $n>1$ are given by
\begin{equation}
U_{n}(\theta) = 
\begin{cases}
    \frac{1}{(\Delta\theta)^2} \sqrt{\frac{2n + 1}{2}}P_{n}\Big(\frac{2(\theta - \Bar{\theta})}{\Delta\theta}\Big) & \text{if $\theta_{\rm min}\leq\theta\leq\theta_{\rm max}$}\\
    0 & \text{otherwise},
  \end{cases}
\label{eq:Un}
\end{equation}
where $\tb=(\theta_{\rm min}+\theta_{\rm max})/2$ and $\Delta\theta = \theta_{\rm max}-\theta_{\rm min}$. We can calculate the $Q_n(\theta)$ by inserting \Eqs\eqref{eq:Un} and \eqref{eq:U1} into \Eqt\eqref{eq:Q_Filter_Function_integral}. We calculated the analytic solution for $Q_1(\theta)$
\begin{equation}
Q_1(\theta)= \frac{2}{\theta^2\Delta\theta\sqrt{2\Delta\theta^2+24\theta^2}}[A(\theta)-A(\theta_{\rm min})]- U_1(\theta) \;,
\label{eq:Q1}
\end{equation}
where
\begin{equation}
A(\theta)=\theta^2\left[\frac{4\theta\tb-6\tb^2}{\Delta\theta}-0.5\right]\;.
\end{equation}
For $n>1$ we found,
\begin{align}
\label{eq:Qn}
  	Q_{n}(\theta) & = \frac{\sqrt{2(2n + 1)}}{(\theta\; \Delta \theta)^{2}} 		\sum\limits_{m=0}^{M} \frac{(-1)^{m}(2n-2m)!}{2^{n}\; m!(n-m)!(n-2m)!} \Big(\frac{2}{\Delta \theta}\Big)^{n-2m} \\ \nonumber
	& \times \left[\frac{(\theta - \Bar{\theta})^{n-2m+1}}{n-2m+1} \Big(\theta - \frac{ \theta - \Bar{\theta}}{n-2m+2}\Big)\right]_{\theta_{\rm min}}^{\theta}  - \ U_{n}(\theta) \nonumber\;,
\end{align}
where $M$ is the floor of $n/2$, $M=\lfloor n/2 \rfloor$ and $Q$ is zero if  $\theta<\theta_{\rm min}$ or $\theta>\theta_{\rm max}$.

\begin{figure}
	\includegraphics[width=\hsize]{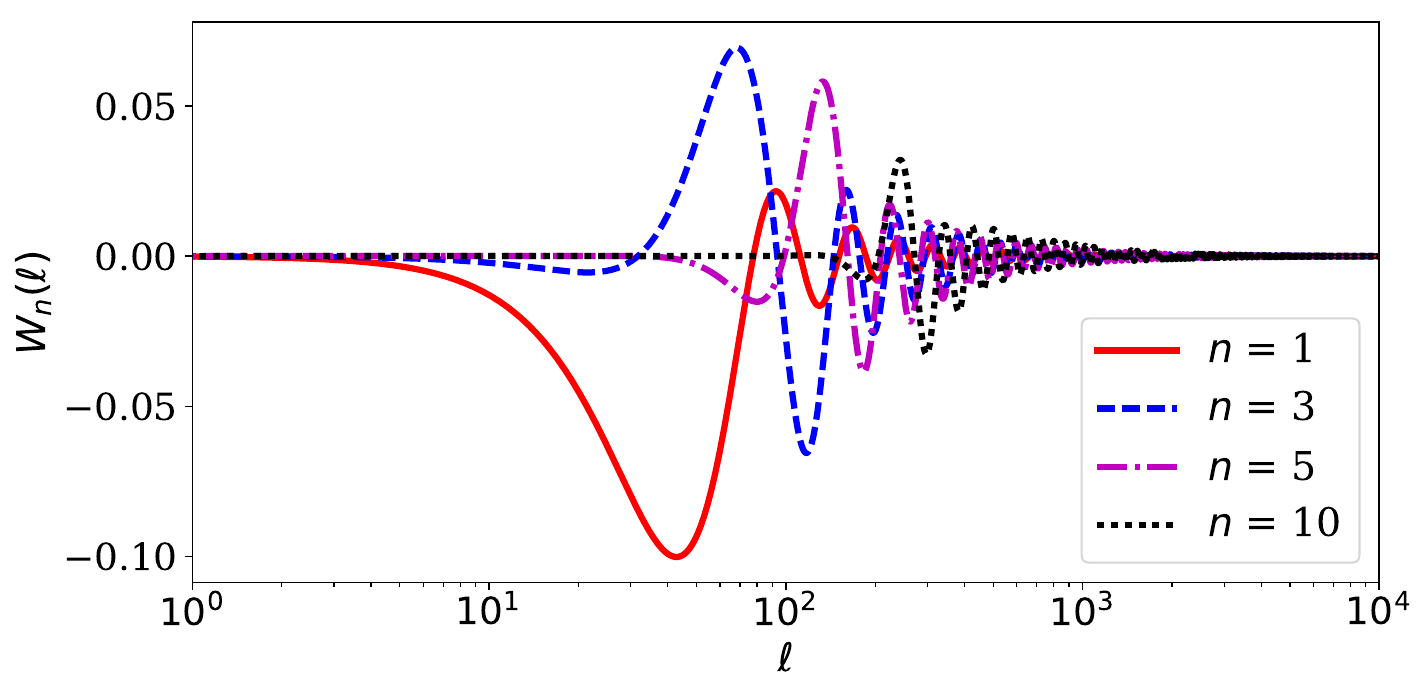}
\caption{The weight function $W_{n}$ for the angular range $[0.5',300']$. Each line depicts a different mode, $n$ (see \Eqt\ref{eq:W_Filter}).}
 \label{fig:WFilters}
\end{figure}

\fig\ref{fig:UQfilters} shows the $U_n(\theta)$ and $Q_n(\theta)$ filters\footnote{There is an error in the calculation of $Q_n(\theta)$ in \cite{Buddendiek} as shown in their Fig.\ts 2. We note that at $\theta_{\rm min}$ and $\theta_{\rm max}$, $Q_n(\theta)=-U_n(\theta)$ since $U_n(\theta)$ is compensated. } for $n=1,3,5$ and 10. All filter functions have the same form irrespective of the angular range used, as indicated by the lower horizontal axis. For reference, we have also included a $\theta$ axis for the angular range of $[0.5',300']$. Both filter functions include more oscillations as $n$ increases, becoming more sensitive to smaller scale variations in the correlation functions. We note that this sensitivity is not limited to small $\theta$-scales, as the oscillations are relatively equidistant in the range of support of the filters. 

To measure $\Psi$ it is more convenient to use their relation with the real space correlation functions (Eqs.\ts\ref{eq:psi_gg_data} and \ref{eq:psi_gm_data}) and use fine binning in $\theta$ to evaluate the integrals. We can increase the accuracy with which $\Psi$-statistics are measured by increasing the number of $\theta$-bins, as long as there is at least one pair of galaxies in each bin. In the case of $\Upsilon$-statistics where the filter functions $U(\theta)$ and $Q(\theta)$ are reduced to Dirac delta functions, fine binning can produce a very noisy $\gamma_{\rm t}(\theta_{\rm min})$ while a broad binning produces an inaccurate estimate of $\gamma_{\rm t}(\theta_{\rm min})$ albeit with smaller errors. Therefore, the measured $\Upsilon(\theta,\theta_{\rm min})$ will either be very noisy or biased as the value of $\gamma_{\rm t}(\theta_{\rm min})$ affects all measurements (see \Eqt\ref{eq:OldUpsilon}).

To produce theoretical predictions for $\Psi$-statistics we use their relation to the angular power spectra, 
\begin{equation}     
\Psi^{\gg}_{n} =\int_{0}^{\infty} \frac{\ell\ \d\ell}{2\pi}  W_{n}(\ell)\ C^{\gg}(\ell)\;,
\label{eq:psi_gg_theory}
\end{equation}
and
\begin{equation}
\Psi^{\gm}_{n} =   \int_{0}^{\infty} \frac{\ell\ \d\ell}{2\pi} W_{n}(\ell)\ C^{\gm}(\ell)\;,
\label{eq:psi_gm_theory}
\end{equation}
where the weight function, $W_{n}$, is
\begin{equation}
W_{n} (\ell) = \int_{\theta_{\rm min}}^{\theta_{\rm max}} \d\theta \theta \ U_{n}(\theta) {\rm J}_{0}(\ell\theta)  = \int_{\theta_{\rm min}}^{\theta_{\rm max}} \d\theta \theta \ Q_{n}(\theta) {\rm J}_{2}(\ell\theta) \;.
\label{eq:W_Filter}
\end{equation}
\fig\ref{fig:WFilters} shows the weight functions, $W_{n} (\ell)$, for the angular range of $\theta\in[0.5',300']$. We show the same modes as in \fig\ref{fig:UQfilters}. The higher modes have an increased weight for larger $\ell$-scales. In general, we see that the $W_{n} (\ell)$ have a compact range of support limiting the information content and hence modelling uncertainties arising from very large and very small $\ell$-scales.


%% file: Surveys.tex
\label{sec:surveys_cosmo}

Throughout this paper we assume flat $\Lambda \rm{CDM}$ models. In the following section, we describe the cosmological pipeline to obtain the theoretical predictions for $C^{\gg,\gm}(\ell)$, $\Psi^{\gg,\gm}_{n}$, $\omega(\theta)$, and $\gamma_{t}(\theta)$. We perform our analysis on two survey setups based on the combined probe analysis of the DES-Y1 data scaled to the full DES area and the combination of KiDS-1000 and BOSS data, described in \sect\ref{sec:surveys}.

\subsection{Cosmological model and pipeline}
\label{sec:cosmological_prediction}

We use the modular cosmological code, {\sc CosmoSIS} \citep{cosmosis} for our predictions, adding  a new $\Psi$-statistics calculation module. The linear matter power spectrum is estimated using {\sc camb}  \citep{CAMB} and its non-linear evolution is calculated using the halo model of \cite{mead/etal:2015}. Galaxy bias is then applied to the matter power spectrum, where applicable.
We then extrapolate the resulting power spectra to $k=500\; h\,$Mpc$^{-1}$ before projecting them using an extended Limber approximation according to \Eqs\eqref{eq:P_gg} and \eqref{eq:P_gm}. The real space correlation functions are calculated using \Eqs\eqref{eq:clustering} and \eqref{eq:ggl}. We follow the integration method described in \cite{asgari/etal:2012} for estimating $\Psi$-statistics using \Eqs\eqref{eq:psi_gg_theory} and \eqref{eq:psi_gm_theory}. 

For our fiducial cosmology, we set the standard deviation of perturbations in a sphere of radius 8\,Mpc$\,h^{-1}$ today, $\sigma_8 = 0.826$, the matter density parameter, $\Om=0.2905$, the baryon density parameter, $\Omega_{\rm b} = 0.0473$, the dimensionless Hubble parameter, $h=0.6898$ (relative to $100\,{\rm km}\,{\rm s}^{-1}\,{\rm Mpc}^{-1}$), the spectral index of the primordial power spectrum, $n_{\rm s}=0.969$, and ignore baryon feedback by setting the $A_{\rm bar}$ parameter for the halo model to $3.13$.

\subsection{Survey setups}
\label{sec:surveys}

\begin{table}
\centering
\caption{\small{Values for the two setups: KiDS-BOSS-like and DES-Y1-5000 \citep[taken from][]{kuijken/etal:2019,DES}. Rows 2-4 show the area in deg$^2$ for the galaxy clustering, GGL and cosmic shear surveys. The fifth row shows the number of source bins where shear is measured for each survey and the following five rows show the number density of galaxies per arcmin$^2$ in each redshift bin starting from the lowest bin. Row eleven  shows the number of lens redshift bins where the position of the galaxies is measured, followed by five rows showing the number density of galaxies per arcmin$^2$  in these redshift bins, which are also ordered from the lowest bin to the highest.}}
\label{tab:setups}
\renewcommand{\arraystretch}{1.2}
\begin{tabular}{ c  c  c  }
                                     & KiDS-BOSS-like           &   DES-Y1-5000   \\ \hline\hline
Clustering Area               &  $9329$              &  $5000$       \\ \hline
GGL Area                       &   $408$             &  $5000$    \\ \hline
Cosmic shear Area           &    $773$             & $5000$ \\ \hline\hline
Number of Source bins      &     5                   & 4 \\ \hline
$\bar{n}_{\rm source\;  1}$    &    0.8            & 1.5 \\ \hline
$\bar{n}_{\rm source\;  2}$    &     1.33        & 1.5 \\ \hline
$\bar{n}_{\rm source\;  3}$    &      2.35        & 1.6 \\ \hline
$\bar{n}_{\rm source\;  4}$   &       1.55       & 0.8 \\ \hline
$\bar{n}_{\rm source\; 5}$     &     1.44          & $-$ \\ \hline\hline
Number of lens bins        &       2                   & 5   \\ \hline
$\bar{n}_{\rm lens\; 1}$         &        0.015           & 0.013 \\ \hline
$\bar{n}_{\rm lens\; 2}$         &        0.015          &  0.034\\ \hline
$\bar{n}_{\rm lens\; 3}$         &         $-$             & 0.051 \\ \hline
$\bar{n}_{\rm lens\; 4}$         &         $-$             & 0.030 \\ \hline
$\bar{n}_{\rm lens\; 5}$         &         $-$             & 0.009 \\ \hline
\end{tabular}
\end{table}

\begin{figure*}
\begin{center}
     \begin{tabular}{c}
	\includegraphics[width=0.5\textwidth]{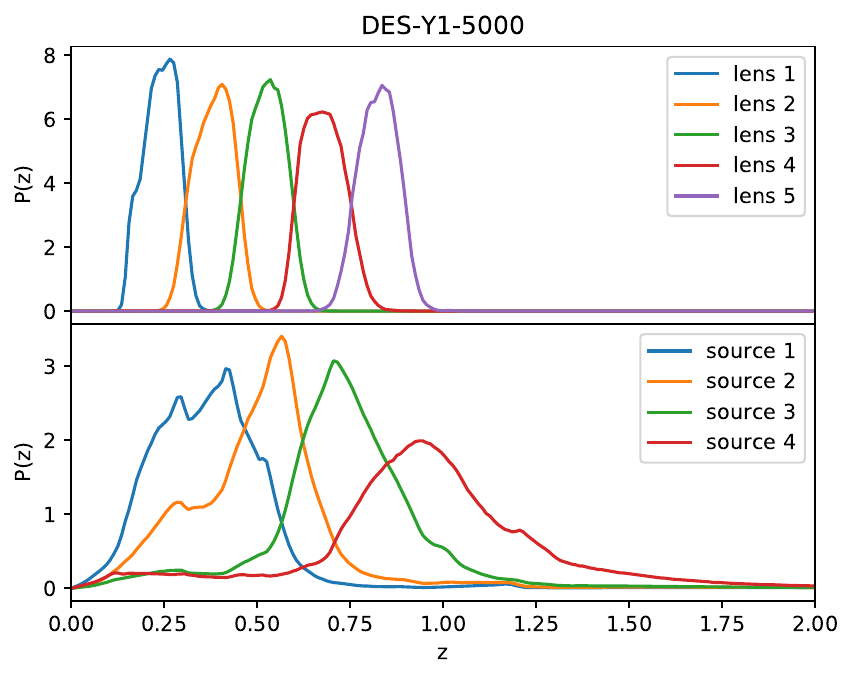}
	\includegraphics[width=0.5\textwidth]{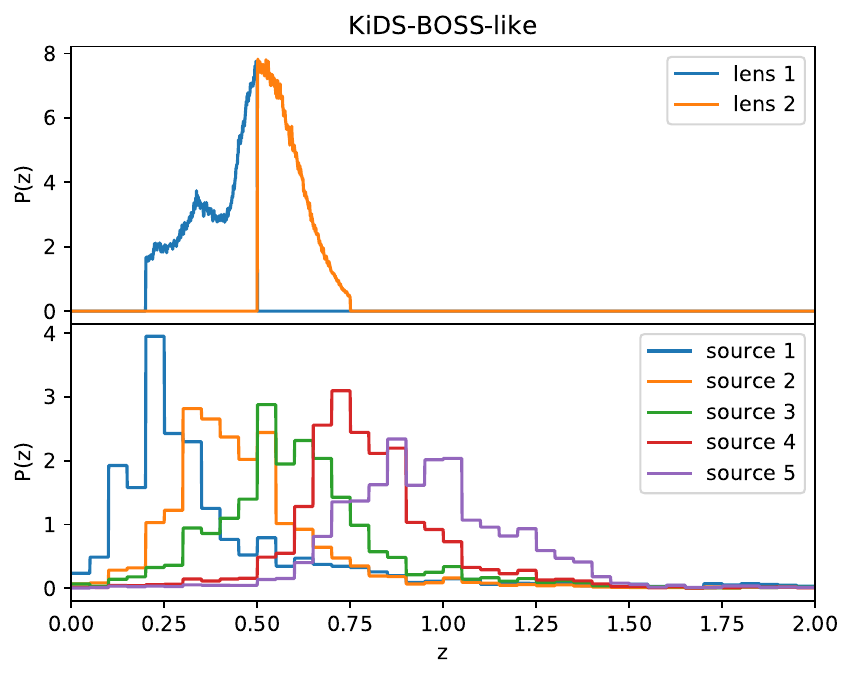}
    \end{tabular}
    \end{center}
    \caption{Normalised redshift distributions for KiDS-BOSS-like and DES-Y1-5000 setups (see \sect\ref{sec:surveys} and \tab\ref{tab:setups}). The top panels show the lens samples, where the galaxy clustering signal is also measured, while the bottom panels show the source distributions used for estimating the GGL signals.}
    \label{fig:nz}
\end{figure*}

The first setup, “KiDS-BOSS-like”, is a 1000 deg$^2$ KiDS-like weak lensing survey combined with a 10,000 deg$^2$ BOSS-like spectroscopic survey.   The second, “DES-Y1-5000”, is a 5000 deg$^2$ DES-like weak lensing survey with an overlapping photometric luminous red galaxy sample. \tab\ref{tab:setups} shows the area, number of redshift bins and number density of galaxies in each redshift bin for these survey setups. To find the covariance for each survey we use these values and the reported ellipticity dispersion for each survey. For the KiDS-BOSS-like case we neglect the cross-covariance terms between the galaxy clustering and GGL signals, since the BOSS area is much larger than KiDS  and their cross-correlation has a negligible effect on the parameter estimation \citep[see][]{joachimi/etal:2020}. The full DES data will likely be deeper with a higher number density of galaxies especially for the highest redshift bins.

The redshift distributions for the KiDS-BOSS-like and DES-Y1-5000 cases are shown in \fig\ref{fig:nz}. The redshift distributions are based on the currently public data of each survey. We use the auto-correlation between the position of the lens galaxies to estimate the galaxy clustering signals and their cross-correlation with the shear of the source galaxies to predict the GGL signals. We use all bin combinations for this analysis although a cross-correlation between a low source bin and high lens bin results in a very low signal-to-noise ratio.

%% file: sensitivity.tex
\label{sec:sensitivity_Fisher}

In this section we quantify the impact of ignoring the scale dependence of galaxy bias on a cosmological analysis combining galaxy clustering and GGL. Here we examine the sensitivity of the two-point statistics described in \sect\ref{sec:Methods} to the differences between galaxy bias models. In \sect\ref{sec:sensitivity} we compare the predicted two-point functions for each model to their expected measurement errors. For this comparison we chose a clustering signal that is calculated for the auto-correlation of the first BOSS lens bin and a GGL signal corresponding to the cross-correlation of the first BOSS lens bin and the highest KiDS source bin (see \fig\ref{fig:nz}). In \sect\ref{sec:Fisher} we include the full survey setups introduced in \sect\ref{sec:surveys} and propagate the errors to cosmological parameters using a Fisher analysis. We also apply scale cuts on all two-point statistics to reduce the biases arising from small scales.

\begin{figure*}
	\includegraphics[width=\textwidth]{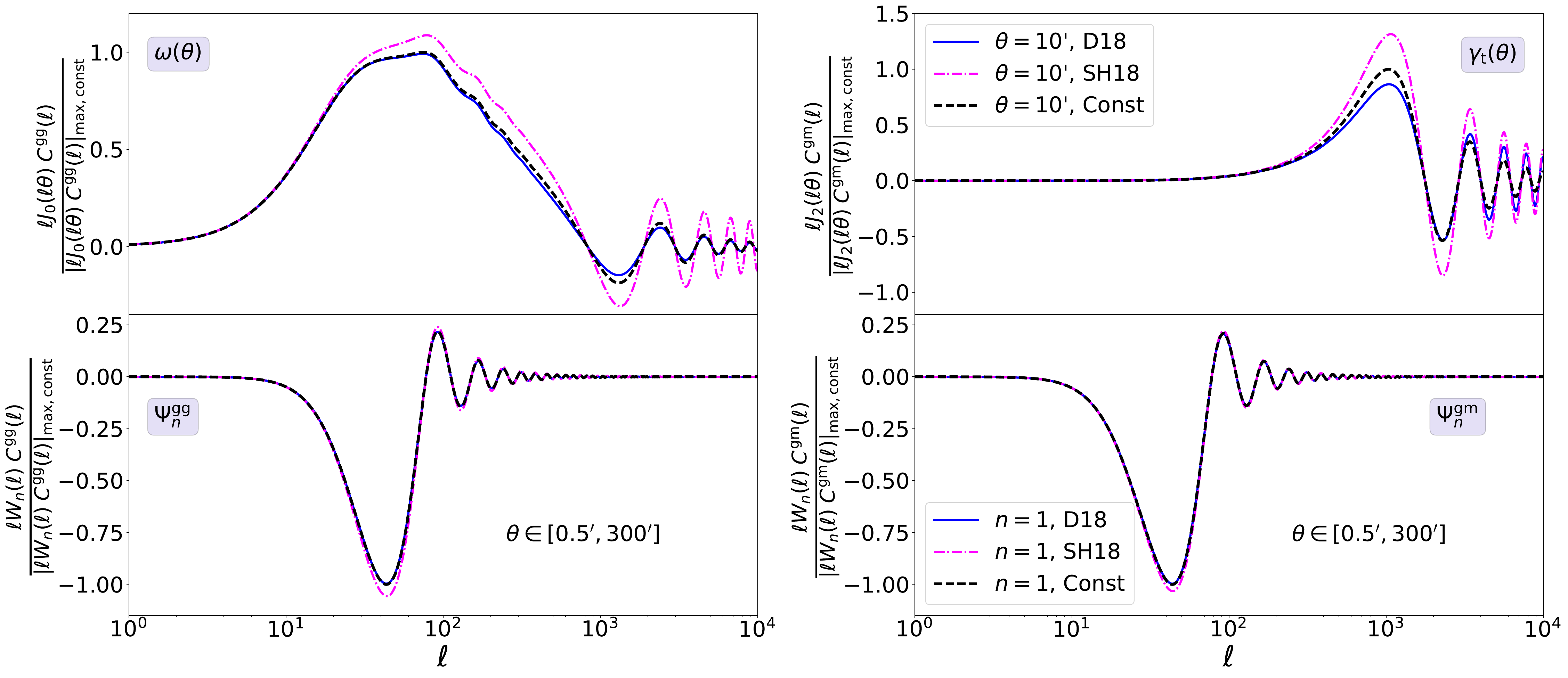}
    \caption{\small{Integrands of correlation functions and $\Psi$-statistics using constant and non-linear galaxy bias models. \textbf{Upper panels: }$\omega(\theta)$ (left) and $\gamma_{\rm t}(\theta)$ (right) for $\theta=10'$.  \textbf{Lower panels: } $\Psi^{\gg}_{n}$ (left) and  $\Psi^{\gm}_{n}$ (right) for  $n$-mode=1 and and angular range of $[0.5',300']$. All integrands are normalised by their absolute maximum value for the constant bias case.}}
    \label{fig:integrands}
\end{figure*}

\begin{figure*}
    \centering
    \includegraphics[width=\textwidth]{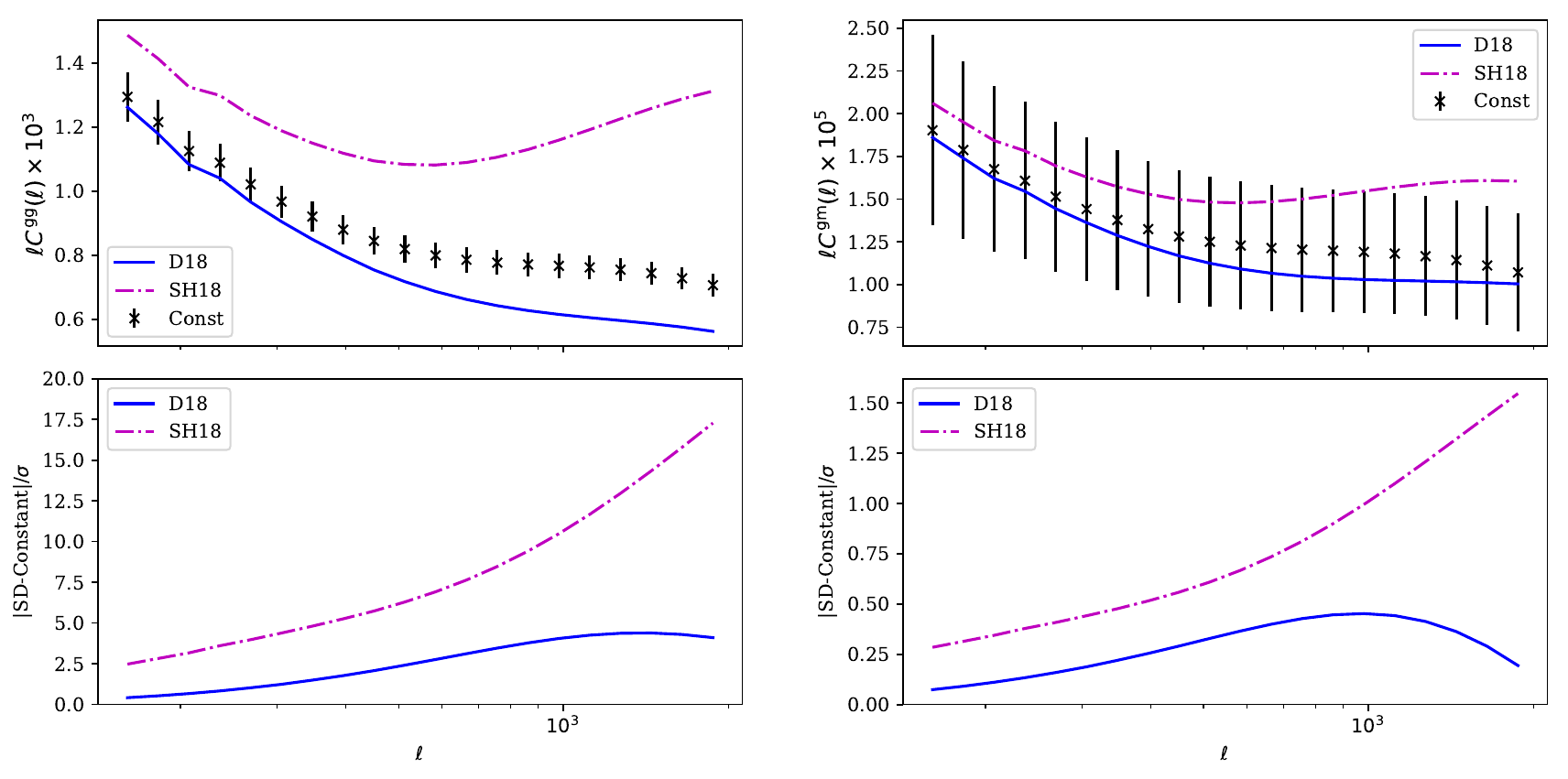}%
    \caption{The angular power spectra (\Eqs\ref{eq:P_gg} and \ref{eq:P_gm}) for the constant bias (black crosses) and the two scale-dependent bias models; SH18 (magenta dot-dashed) and D18 (blue solid). 
    We use the KiDS-BOSS-like setup here and show the clustering signal for the autocorrelation of redshift bin 1 (left) and the GGL signal for lens bin 1 and source bin 5 (right).
     The lower panels show the ratio of the absolute difference between the scale-dependent (SD) and constant bias models over the error calculated for the constant bias model. The $\ell$-range shown here corresponds to the scales used in \citet{vanuitert/etal:2018}.  }
    \label{fig:Fourier_space_sensitivity}
\end{figure*}

\begin{figure*}
    \centering
    \includegraphics[width=\textwidth]{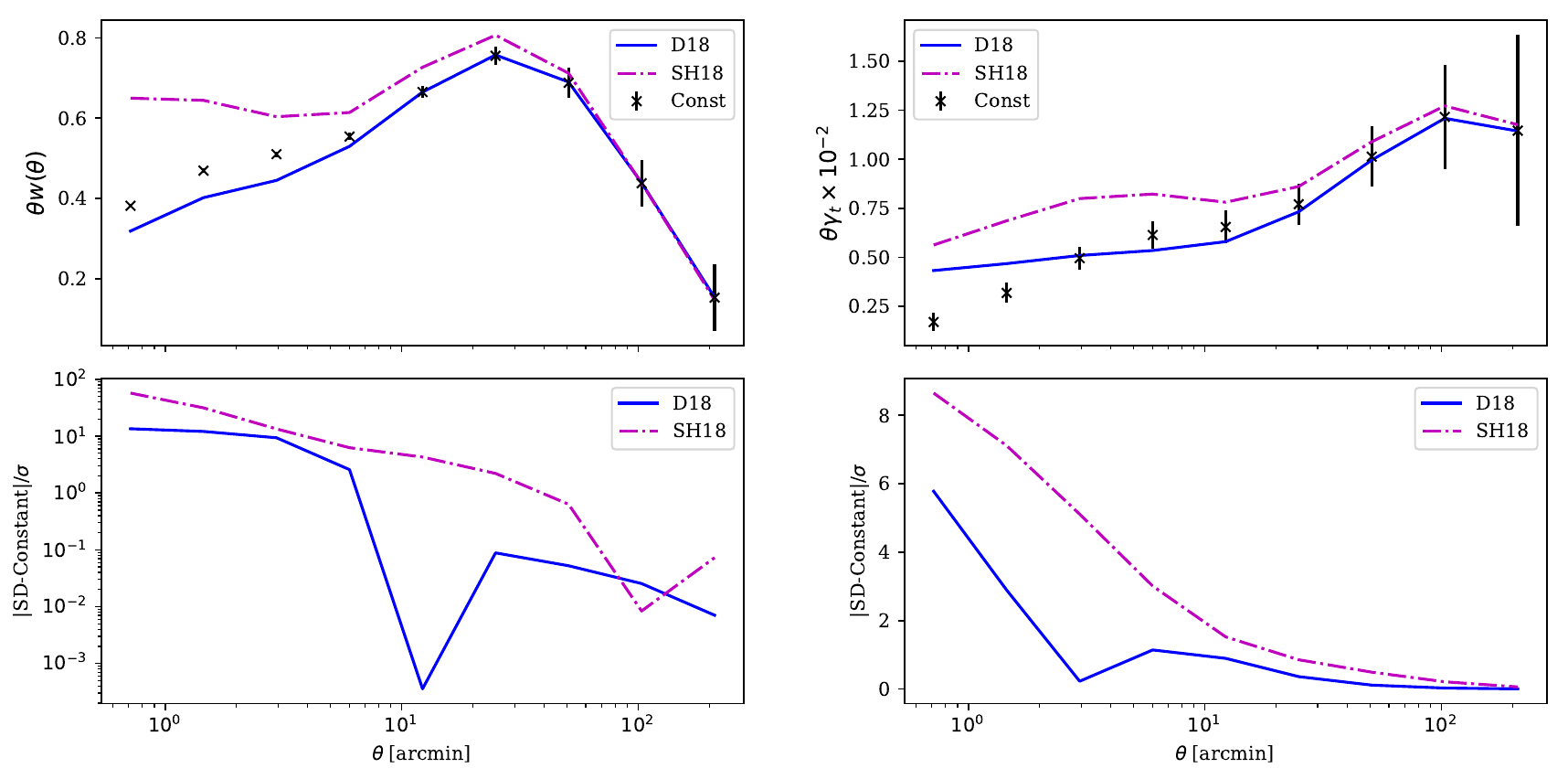}%
    \caption{Angular two-point correlation functions (\Eqs\ref{eq:clustering} and \ref{eq:ggl}) for constant bias (black crosses), SH18 (magenta dot-dashed) and D18 (blue solid). The angular clustering of lens galaxies in the first bin is shown on the left and on the right we show results for GGL using lens bin 1 and source bin 5, with the KiDS-BOSS-like setup. The lower panels show the absolute signal-to-noise difference between constant and scale-dependent (SD) models. }
   \label{fig:Real_space_sensitivity}
\end{figure*}

\begin{figure*}
    \centering
    \includegraphics[width=\textwidth]{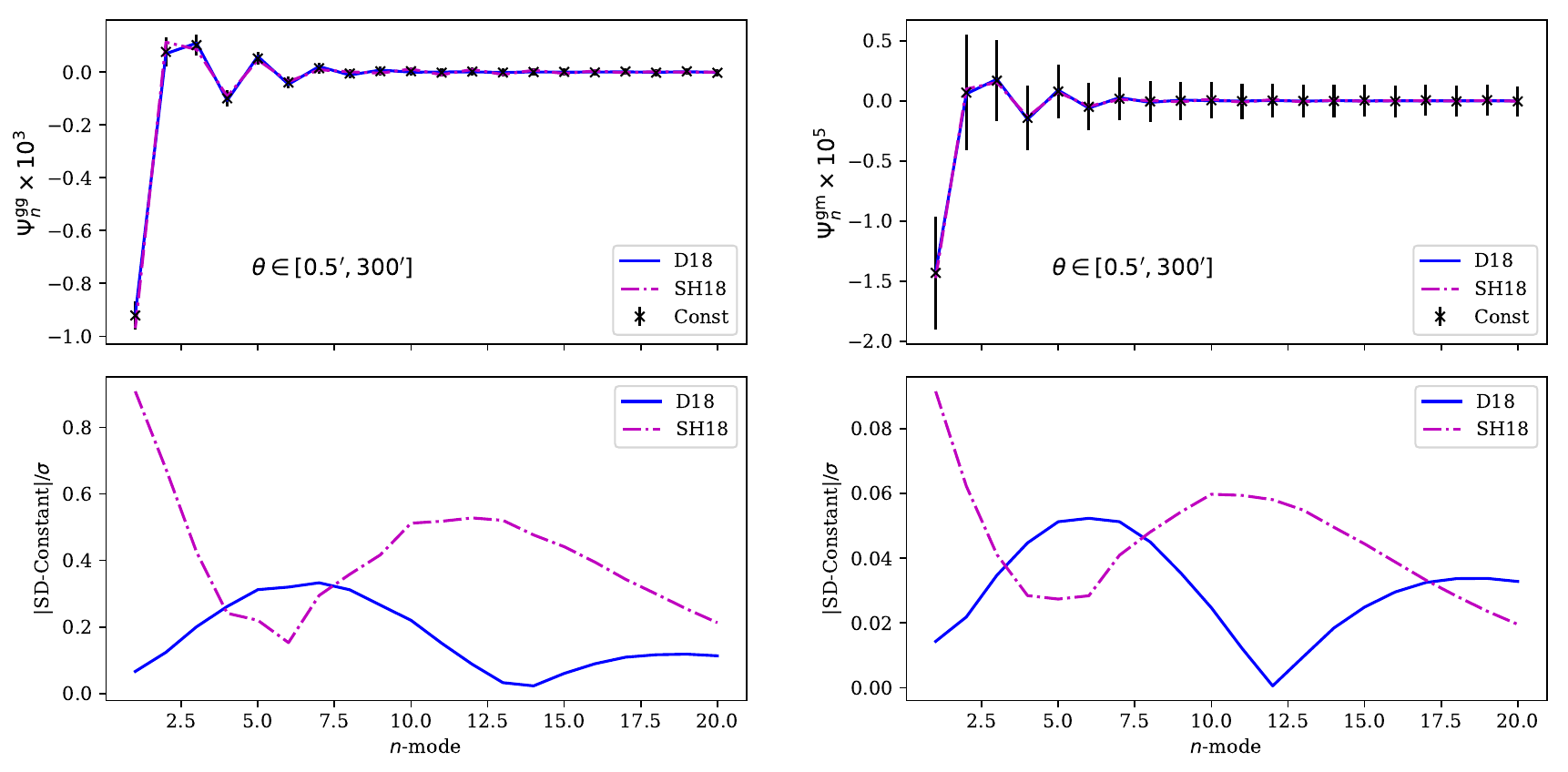}%
    \caption{ $\Psi$-statistics (\Eqs\ref{eq:psi_gg_theory} and \ref{eq:psi_gm_theory}) for constant bias (black crosses), SH18 (magenta dot-dashed) and D18 (blue solid), defined for $0.5'\leq\theta\leq300'$. The clustering (left) and GGL (right) results are shown for the autocorrelation of the first lens bin and its cross-correlation with the fifth source bin, respectively. We used the KiDS-BOSS-like setup here. In the lower panels we show the absolute signal-to-noise difference between constant and scale-dependent (SD) models.}
   \label{fig:psi_compare}%
\end{figure*}

\subsection{Sensitivity of statistics to galaxy bias}
\label{sec:sensitivity}

The three two-point statistics introduced in \sect\ref{sec:Methods} have different $\ell$-scale dependences. The $\ell$ dependence of $C(\ell)$ is trivial but the other two families have a more complex weighting per $\ell$-mode. 

\fig\ref{fig:integrands} shows the integrands of the real space correlation functions (upper panels) and the $\Psi$-statistics (lower panels), where we compare them for the constant bias (black dashed), the SH18 (magenta dot-dashed) and D18 (blue solid) bias models. All integrands are normalised by the absolute maximum value of their constant bias case. The integrands of the correlation functions are shown for $\theta=10'$, while the $\Psi$-statistics are defined over the angular range of $[0.5',300']$ and shown for $n=1$.
Comparing the panels we see that the $\ell$-range used to estimate $\Psi_n$ is more concentrated than that of the real space correlation functions and as a result we expect them to be less sensitive to non-linear galaxy bias when defined over the same angular range. We also see that the SH18 model shows a more prominent difference to constant bias compared to the D18 model. An extended version of this figure can be found in \fig\ref{fig:integrands_app}, where we show that our conclusions still hold in the case of larger theta scales or $n$-modes, since the Bessel functions have a wider range of support compared to the $\Psi$-statistics weight functions.

In \figs\ref{fig:Fourier_space_sensitivity}, \ref{fig:Real_space_sensitivity} and \ref{fig:psi_compare} we show the theoretical predictions for the angular power spectra, real-space correlation functions and $\Psi$-statistics, respectively. The upper panels show the signal for constant bias (crosses with errorbars), SH18 model (magenta dot-dashed) and D18 (blue solid). The errorbars are calculated from theory using the values discussed in \sect\ref{sec:surveys} and assuming a constant galaxy bias model. The lower panels show the absolute difference in signal-to-noise between the signals from the scale dependent models and the linear bias model. The left hand panels show results for galaxy clustering while the right hand ones belong to GGL.

The angular power spectra in \fig\ref{fig:Fourier_space_sensitivity} are calculated for 20 logarithmic $\ell$-bins between $\ell=150$ and $\ell=2000$. We chose this range of $\ell$ based on the combined probe analysis of \cite{vanuitert/etal:2018}, who divided the data into 5 logarithmic bins instead. The real space correlation functions and $\Psi$-statistics in \figs\ref{fig:Real_space_sensitivity} and \ref{fig:psi_compare} are defined using the angular range of $[0.5', 300']$. The correlation functions are logarithmically binned into 9 $\theta$-bins for this angular range.

The upper panels of \fig\ref{fig:Real_space_sensitivity} show that the three models converge to the constant bias value for large $\theta$-scales. The SH18 model, as can be seen in \fig\ref{fig:bias_models}, starts to show a scale-dependent behaviour at smaller $k$-scales compared to the D18 model, which qualitatively translates to larger $\theta$-scales. In \fig\ref{fig:Fourier_space_sensitivity} we see that even for the lowest $\ell$-modes plotted here, the SH18 model shows differences with the constant bias model. We note that the relations in \Eqs\eqref{eq:P_gg} and \eqref{eq:P_gm} complicate the $k$-dependence of the $C(\ell)$ functions, through the line-of-sight integrals. Comparing the lower panels of \figs\ref{fig:Fourier_space_sensitivity}, \ref{fig:Real_space_sensitivity} and \ref{fig:psi_compare} we see that $\Psi$-statistics are much less sensitive to the scale dependence of these galaxy bias models (up to $0.9\sigma$), compared to the correlation functions (up to $60\sigma$) and angular power spectra (up to $18\sigma$). We also note that the DES-Y1 analysis that used correlation functions chose a variable scale cut depending on the redshift bins considered to minimise the impact of smaller physical scales on their results ($\theta=14'$ to $ 43'$ for $w(\theta)$ and $\theta=21'$ to $64'$ for $\gamma_{\rm t}$).

In general, the expected biases arising from the clustering signal is larger owing to two reasons. Firstly, the galaxy clustering signal depends on $b^2(k)$, while the GGL signal scales with $b(k)r(k)$, where $r(k)$ has a less pronounced scale dependence. Secondly, the signal-to-noise ratio for the GGL signal is much lower than the clustering signal for the KiDS-BOSS-like setup, since the overlap area between KiDS and BOSS is much smaller than the full BOSS area. When considering the DES-Y1-5000 setup we find more similar values for the signal-to-noise ratios of the GGL and clustering signals, given that the overlap area between the lens and source samples is equal to the area for each sample (see \tab\ref{tab:setups}).

\subsection{Error propagation to cosmological parameters}
\label{sec:Fisher}

\begin{figure}
    \centering
    \includegraphics[width=0.45\textwidth]{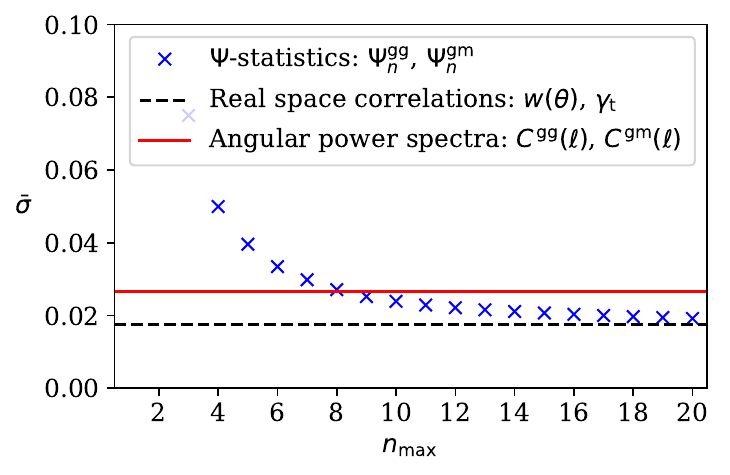}
    \caption{The geometric mean error on the cosmological parameters as a function of the number of modes (see \Eqt\ref{eq:fig_mer}). The survey setup is KiDS-BOSS-like with an angular range of $0.5'\leq\theta\leq300'$. Six free parameters, $\Om$, $\sigma_8$, $n_{\rm s}$ and $\Omega_{\rm b}$,  $h$ and $b_0$ are allowed to vary in this analysis. The horizontal lines show the average error on the parameters using the real space correlation functions (black dashed) measured for the same angular range and the angular power spectra (red solid) for $150\leq\ell\leq2000$. }
    \label{fig:Figure_Of_Merit}
\end{figure}

We use a Fisher formalism to propagate the systematic errors, introduced by neglecting the scale dependence of the galaxy bias, to cosmological parameters. A Fisher analysis provides us with a lower limit to the parameter constraints, which is accurate for parameters with Gaussian distributions. The Fisher matrix is defined as the expectation value of the second order derivative of the log-likelihood, $\mathcal{L}$, of the model given the data with respect to the model parameters, $\phi_{p}$,
\begin{equation}
F_{p q}=\bigg\langle-\frac{\partial^{2}\mathcal{L}}{\partial \phi_{p}\partial\phi_{q}}\bigg\rangle\;.
\end{equation}
If we assume that the data is also Gaussian distributed, then we can skip estimating the likelihood and write the Fisher matrix as,
\begin{equation}
\label{eq:Fisher_expanded}
F_{p q} =\frac{1}{2}{\rm Tr}[\mathbfss{C}^{-1}\mathbfss{C}_{,p}\mathbfss{C}^{-1}\mathbfss{C}_{,q}+\mathbfss{C}^{-1}\mathbfss{M}_{pq}]\;,
\end{equation}
where Tr stands for trace, $\mathbfss{M}_{pq}=\pmb{\mu}_{,p}\pmb{\mu}_{,q}^{T}+\pmb{\mu}_{,q}\pmb{\mu}_{,p}^{T}$ and $\pmb{\mu}_{,p}$ is the derivative of the observable with respect to the model parameter $\phi_{p}$, $\mathbfss{C}$  is the covariance matrix of the data and $\mathbfss{C}_{,p}$ is the derivative of the covariance matrix with respect to $\phi_{p}$ \citep[see for example][]{Tegmark97}. The inverse covariance matrix scales with the area of the survey (see \Eqt\ref{eq:cov_all}), consequently the second term in \Eqt\eqref{eq:Fisher_expanded} will dominate for larger surveys. Therefore, here we do not include the first term in calculating the Fisher matrices. To compute the derivatives we use a five-point stencil method with a step size equal to $2\%$ of the parameter values. This method uses a combination of four points near the fiducial value. 

Before quantifying the effect of the scale-dependent bias models, we compare the information content in $\Psi$-statistics with the two-point statistics discussed in \sect\ref{sec:Methods}. For this task we defined a figure-of-merit that presents us with a measure of the average error on the model parameters. We define this figure-of-merit using the fact that for a Gaussian distributed parameter space, the square root of the determinant of the Fisher matrix is inversely proportional to the volume of the confidence regions. 
Therefore, to estimate the size of the errorbars on parameters we calculate,
\begin{equation}
	\centering
	\bar{\sigma} = \left(\frac{1}{\sqrt{|\mathbfss{F}|}}\right)^{1/P}\; ,
\label{eq:fig_mer}
\end{equation}
where $P$ is the number of free parameters and ${|\mathbfss{F}|}$ is the determinant of the Fisher matrix. For the rest of our analysis we choose to vary five cosmological parameters, $\Om$, $\sigma_8$, $h$, $n_{\rm s}$ and $\Omega_{\rm b}$ as well as a number of effective galaxy bias parameters, $b_{z_i}$, one for each lens redshift bin. Although we have not included any redshift evolution in our galaxy bias modelling, we include these extra bias parameters to account for the fact that the kernels in \Eqs\eqref{eq:P_gg} and \eqref{eq:P_gm} have a redshift dependence, which can in turn produce sensitivity to different parts of the scale-dependent galaxy bias curves in \fig\ref{fig:bias_models}. In addition, the $3\times2$pt analysis of \cite{DES}, \cite{Joudaki_KiDS_2dFLenS} and \cite{vanuitert/etal:2018} all adopted this method.

\fig\ref{fig:Figure_Of_Merit} shows  the geometric mean of error on parameters, $\bar{\sigma}$, as a function of the number of $n$-modes used in the analysis with $\Psi$-statistics starting with $n=1$. 
The horizontal solid and dashed lines show the expected value of $\bar{\sigma}$ when analysing the data using the angular power spectra and the real space correlation functions, respectively. 
All the survey properties correspond to the KiDS-BOSS-like setup (see \sect\ref{sec:surveys} and \tab\ref{tab:setups}), with the angular range of $[0.5', 300']$ for $\Psi$-statistics and correlation functions. The angular power spectra are calculated for 20 logarithmic bins between $\ell=150$ and $\ell=2000$.

$\Psi$-statistics contain all the information in the correlation functions that comply with the compensation condition in \Eqt\eqref{eq:compensated_condition}. But this information is shared between the different modes. \fig\ref{fig:Figure_Of_Merit} shows that as more modes are added to the analysis, the value of $\bar{\sigma}$ from $\Psi$-statistics gets closer to that calculated using the correlation functions. We also see that the first few modes contain most of the information, indicated by the sharp decrease in $\bar{\sigma}$, and the higher modes add incremental information on the parameters. The information content of the angular power spectra for $150\leq\ell\leq 2000$ is lower than the correlation functions for $0.5'\leq\theta\leq 300'$, which means that the average size of the confidence regions for parameters is larger and therefore, $\bar{\sigma}$ is also larger for $C(\ell)$ compared to $w(\theta)$ and $\gamma_{\rm t}$. Hence, we can conclude that $\Psi$-statistics have essentially the same constraining power as the correlation functions, when defined on a wide angular range, but as seen in \figs\ref{fig:Real_space_sensitivity} and \ref{fig:psi_compare} they are significantly less sensitive to the scale-dependence of the galaxy bias.

\begin{figure*}
 \begin{center}
     \begin{tabular}{c}
  \includegraphics[width=0.3\linewidth]{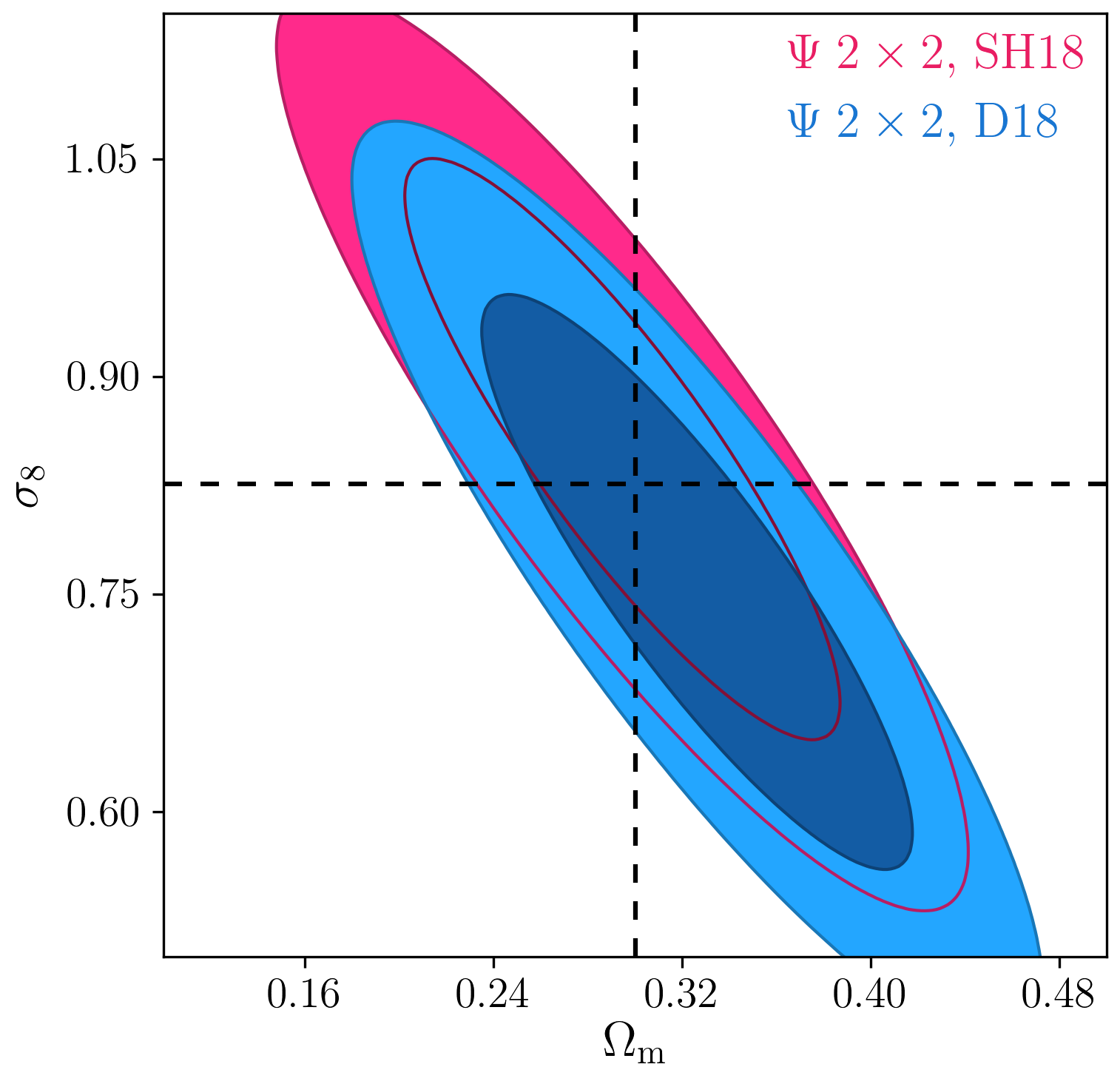}
  \includegraphics[width=0.3\linewidth]{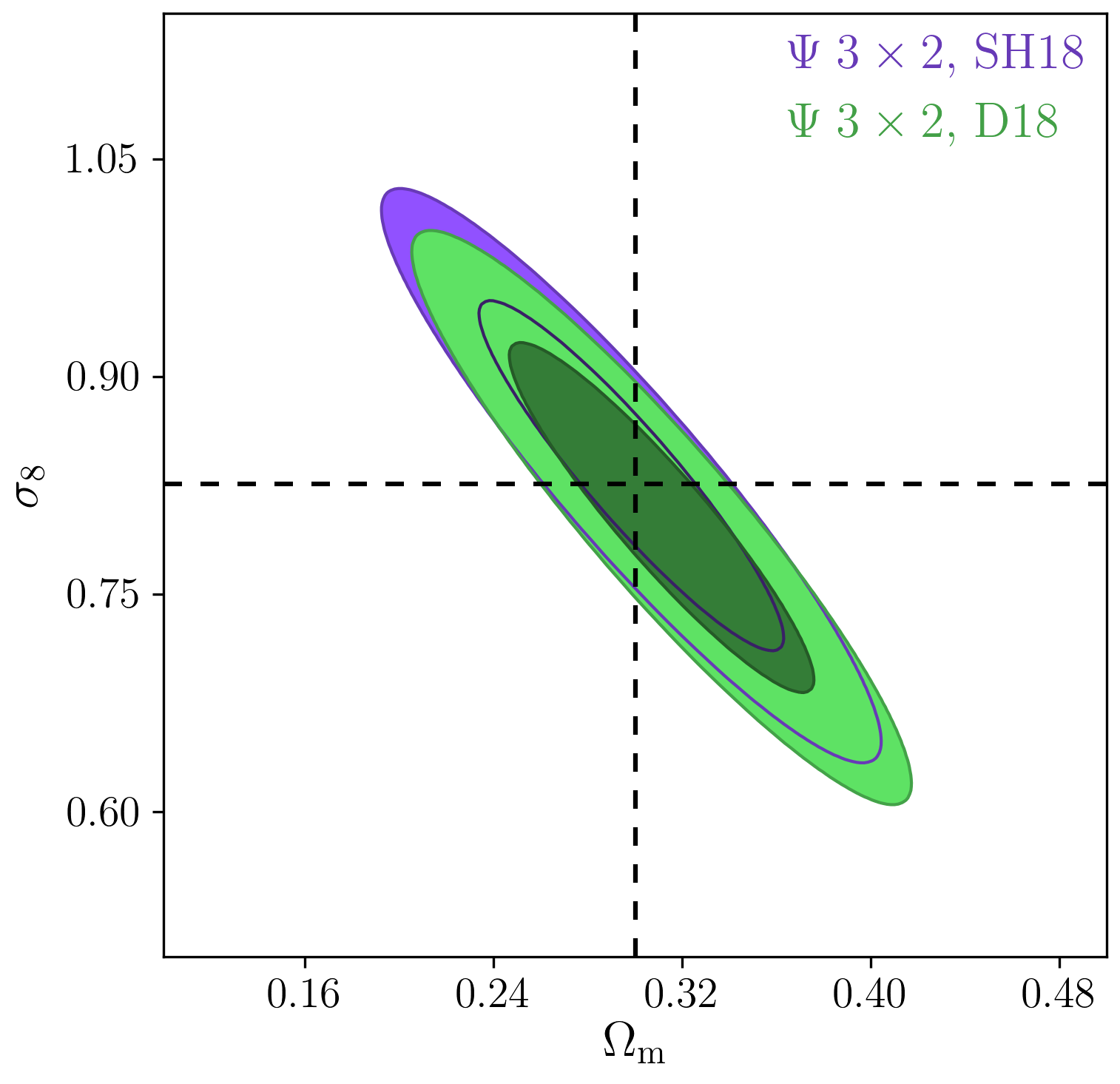}
  \includegraphics[width=0.3\linewidth]{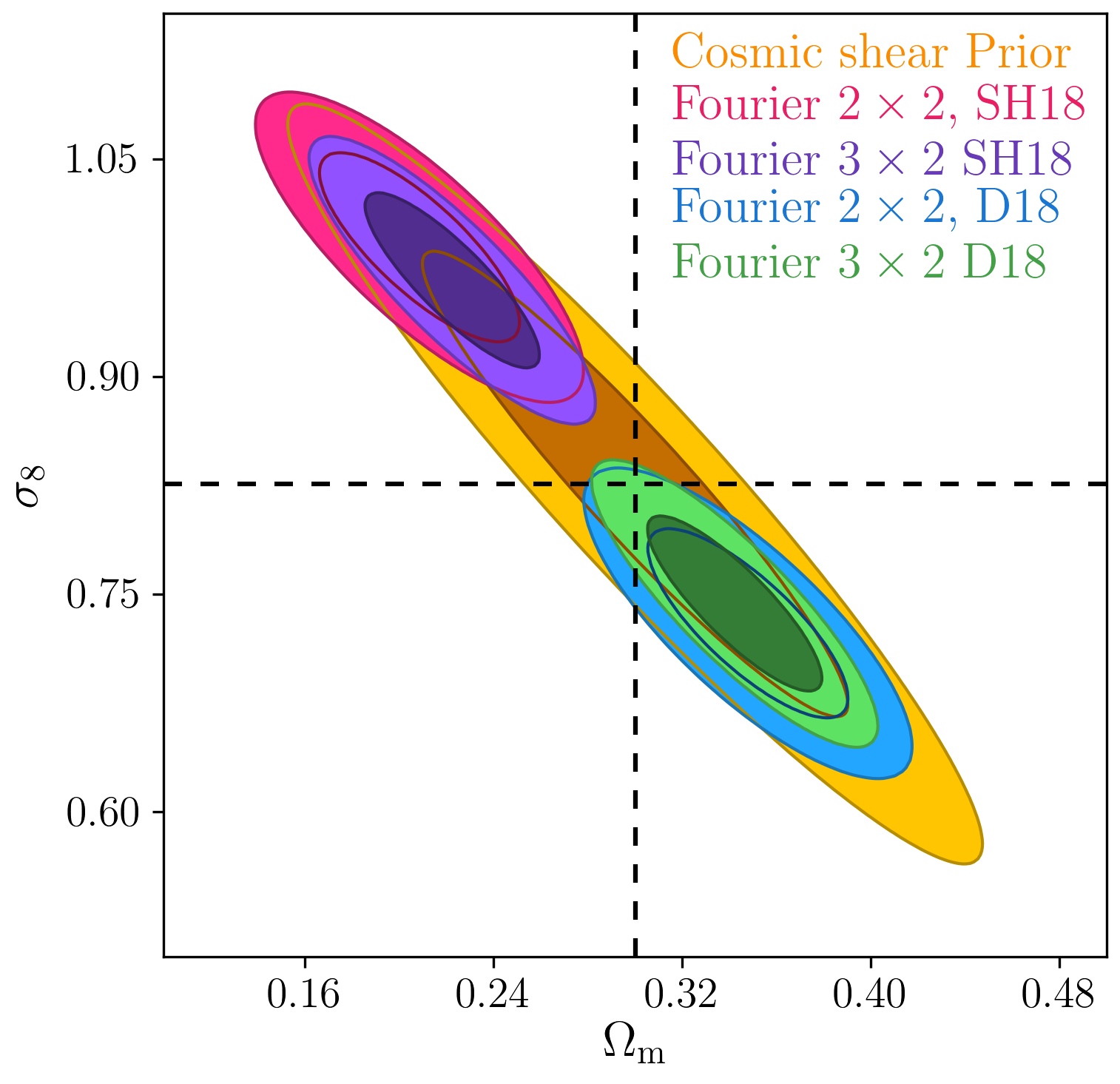}
  \end{tabular}
  \end{center}
\caption{{\small KiDS-BOSS-like expected parameter errors. Mock data is analysed assuming constant galaxy bias. Results are shown for data produced using SH18 and D18 non-linear galaxy bias models (see \sect\ref{sec:Bias_Model}). The input values for $\sigma_8$ and $\Om$ are shown by the dashed lines.  The name tag $2\times 2$, corresponds to the combination of galaxy clustering and galaxy-galaxy lensing. The cosmic shear prior (orange) based on a KiDS-like analysis is combined with the  $2\times 2$ contours to form the $3\times 2$ results. The left hand and the middle panels belong to $\Psi$-statistics defined on $\theta\in[8',300']$, while the right panel shows results for $C(\ell)$ with $150<\ell<1050$. }}
\label{fig:KiDS-BOSS-like_psi}
\end{figure*}

\begin{figure*}
 \begin{center}
     \begin{tabular}{c}
  \includegraphics[width=0.25\linewidth]{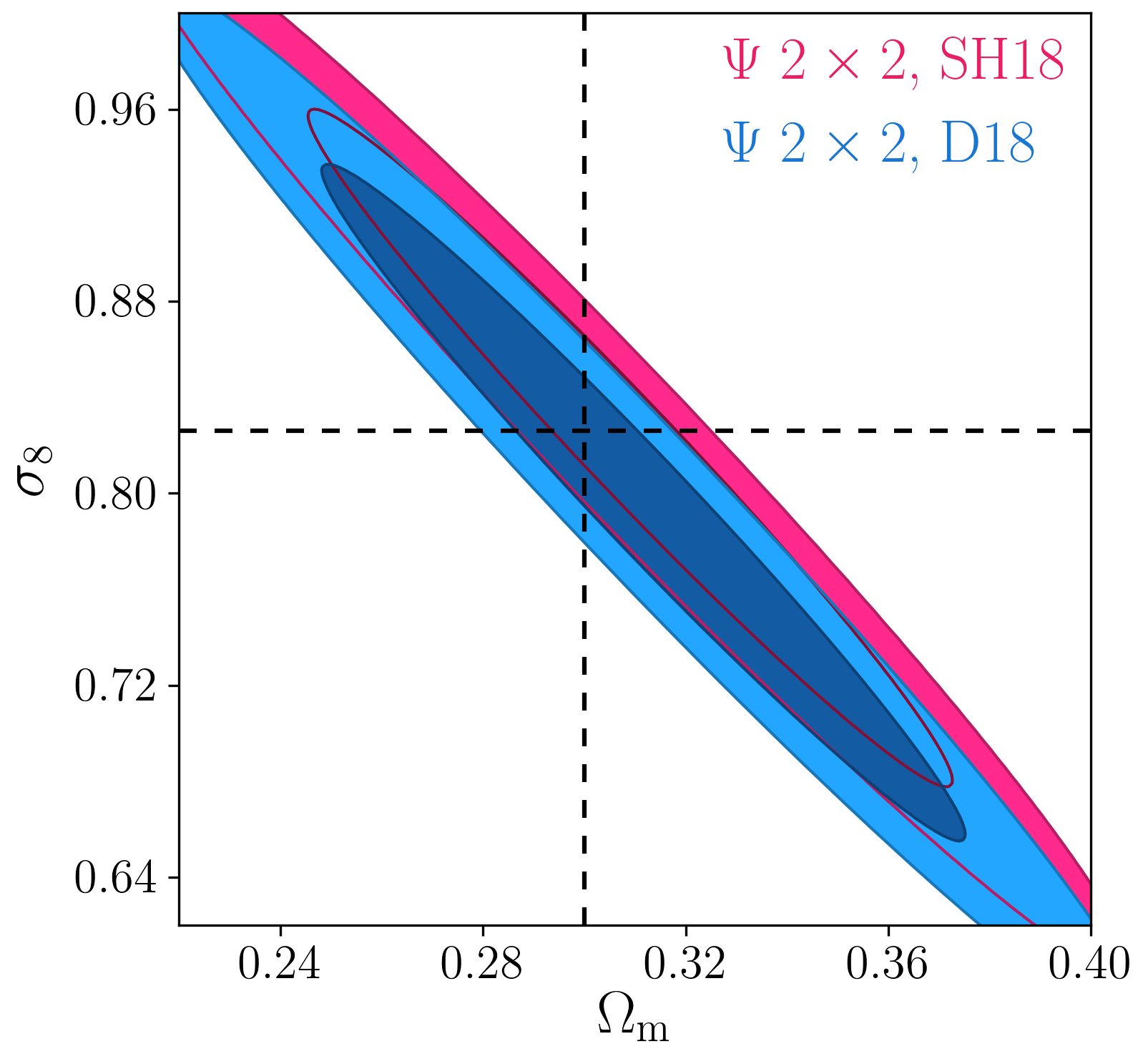}
  \includegraphics[width=0.25\linewidth]{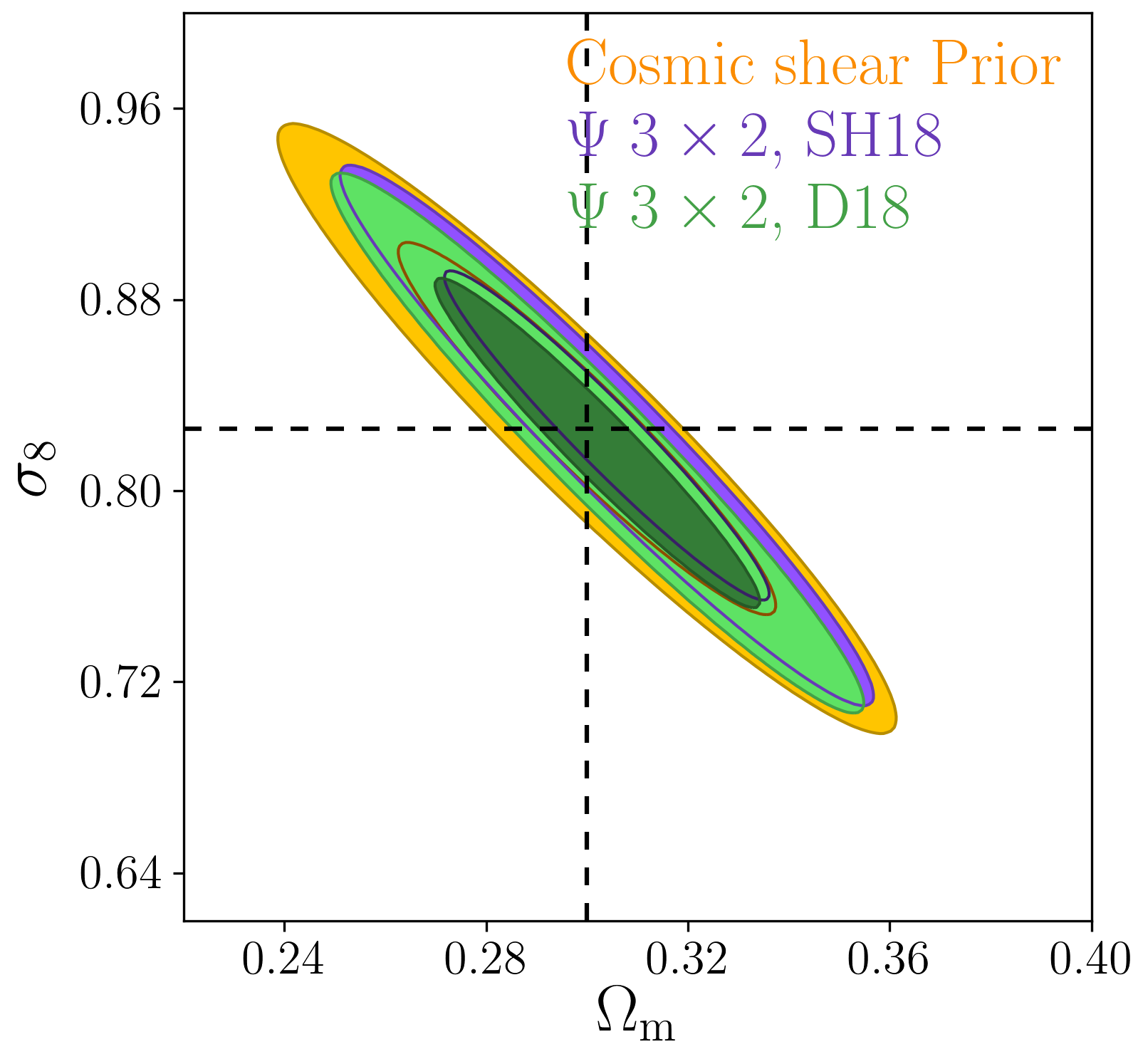}
  \includegraphics[width=0.25\linewidth]{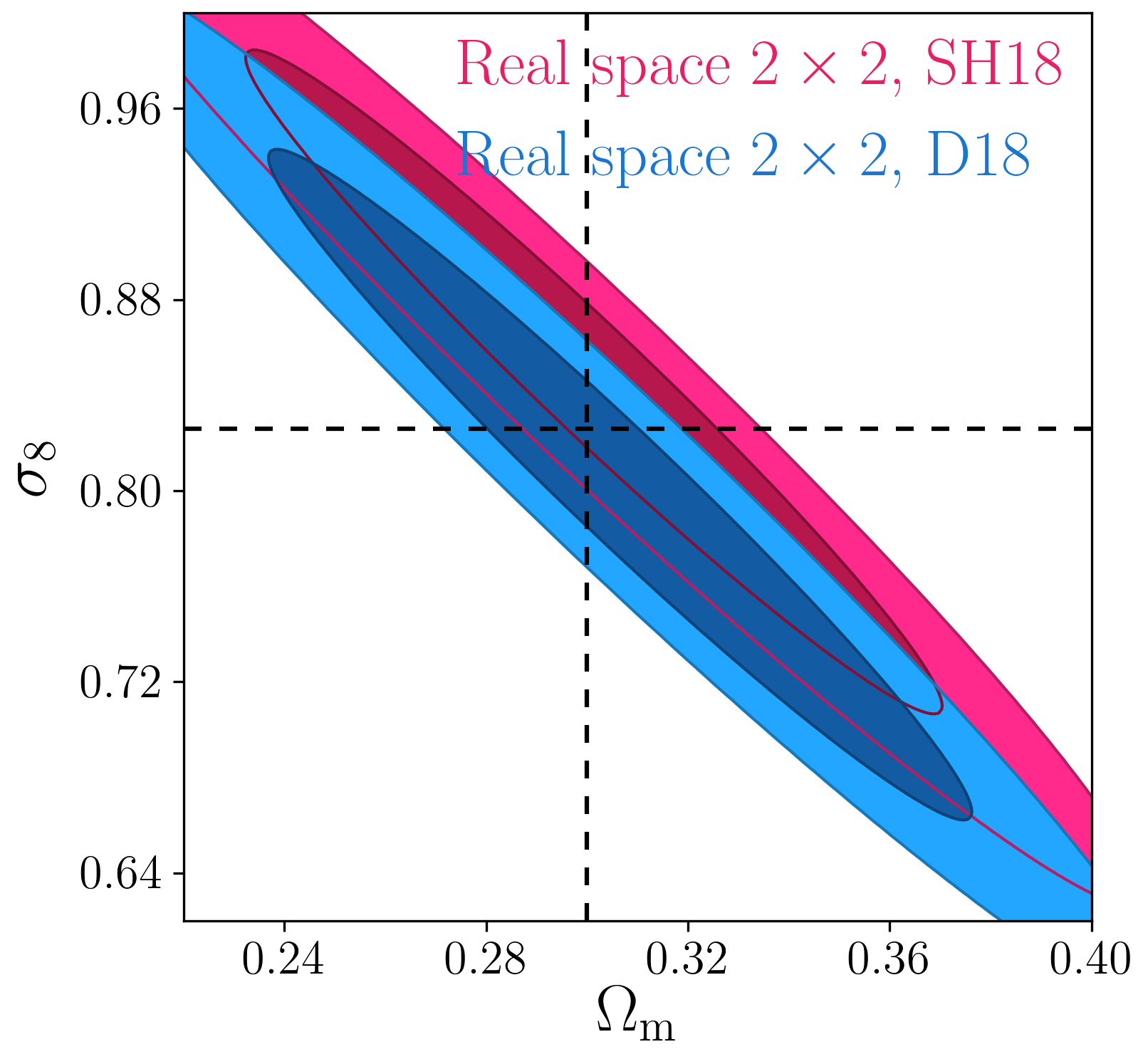}
  \includegraphics[width=0.25\linewidth]{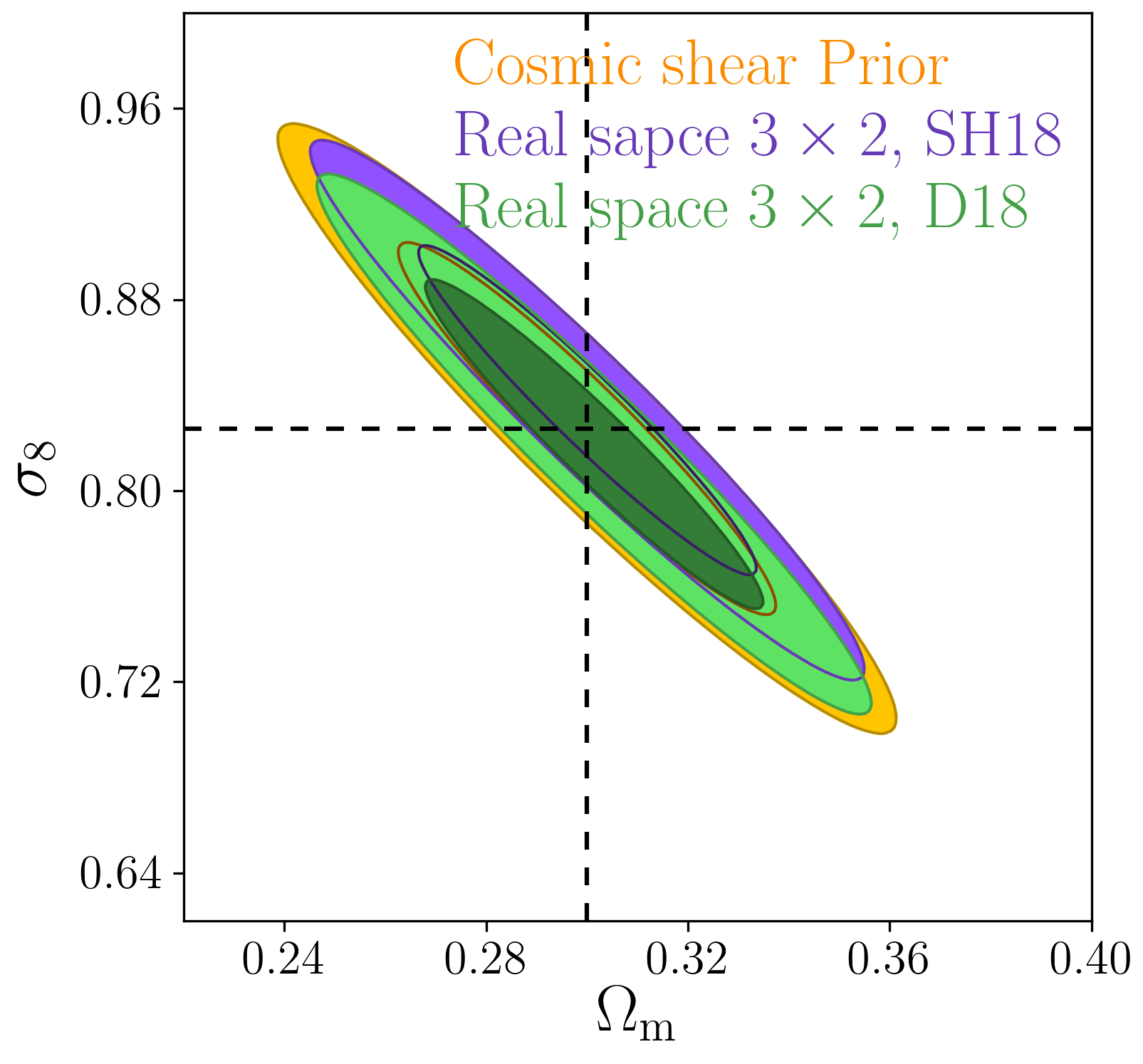}
  \end{tabular}
  \end{center}
\caption{DES-Y1-5000 expected parameter errors. The mock data is produced using non-linear galaxy bias models, SH18 and D18 as described in \sect\ref{sec:Bias_Model}.The analysis, on the other hand, assumes that galaxy bias is constant. The input values for the parameters $\sigma_8$ and $\Om$  are shown with dashed lines. The combination of clustering and GGL is shown with a  $2\times 2$ name tag, while $3\times  2$ denotes the combination of clustering, GGL and the cosmic shear prior based in on a DES-Y1-5000 analysis (see \sect\ref{sec:surveys_cosmo}). The two left hand panels show results for $\Psi$-statistics with  $\theta\in[10',300']$ and $n$-modes$<8$, while the two right hand panels belong to real space correlation functions, $w(\theta)$ and $\gamma_{\rm t}(\theta)$ with the scale cuts used in the DES-Y1 analysis of \citet{DES}. }
\label{fig:DES-Y1-5000_psi}
\end{figure*}

\begin{table*}
\caption{ Error and systematic bias for $\S8$  and   $\Sp$. We define 
$\S8$ and $\Sp$  as the shorter and the longer dimensions of Fisher ellipses, respectively. 
Values are shown for both KiDS-BOSS-like and DES-Y1-5000 setups. In both cases the results are first shown for $\Psi$-statistics. In the case of KiDS-BOSS-like we compare these results with a Fourier analysis, while for DES-Y1-5000 we show results for the real space analysis. The second column shows whether a $2\times2$pt (GGL and clustering) or a $3\times2$pt (GGL and clustering with a cosmic shear prior) is considered for its corresponding row. For this analysis we assume that the data comes from either SH18 or D18 scale-dependant galaxy bias models, but the fitted models assume that galaxy bias is scale-independent.  }
\begin{center}
\begin{adjustbox}{max width=\textwidth}

\bgroup
\def\arraystretch{1.5}
\begin{tabular}{ c c c  c  c  c  c | c  c  c  c  c  }
&&\multicolumn{2}{c}{$\DS8$}&$\sigS8$&\multicolumn{2}{c|}{$\DS8/\sigS8$}&\multicolumn{2}{c}{$\DSp$}&$\sigSp$&\multicolumn{2}{c}{$\DSp/\sigSp$}
\\ \cline{3-12}
& & SH18 & D18 &  & SH18 &  D18  & SH18 &  D18 &  & SH18 & D18
 \\ \hline
  
  
\multirow{2}{*}{KiDS-BOSS-like, $\Psi$} 
& $2\times 2$pt &  0.004 & -0.002 & 0.028 & 0.15 & -0.07 & -0.024 & 0.072 & 0.145 & -0.17 & 0.50
\\ \cline{2-12}
& $3\times 2$pt & 0.002 & -0.001  & 0.014 & 0.12 & -0.09 & -0.005& 0.025 & 0.089 & -0.06 & 0.29 
\\ \cline{1-12}

%

\multirow{2}{*}{KiDS-BOSS-like, $C(\ell)$} 
& $2\times 2$pt & 0.004 & -0.007 & 0.015 & 0.24 & -0.49 & -0.187 & 0.107 & 0.049 & -3.81 & 2.18
\\ \cline{2-12}
 & $3\times 2$pt & 0.002 & -0.004 & 0.011 & 0.18 & -0.37 & -0.161 & 0.093 & 0.045 & -3.55 & 2.06
\\ \hline \hline


\multirow{2}{*}{DES-Y1-5000, $\Psi$} 
& $2\times 2$pt & 0.005 & -0.001 & 0.008 & 0.69 & -0.19 & 0.010 & 0.032 & 0.108 & 0.09 & 0.30
\\ \cline{2-12}
 & $3\times 2$pt & 0.002 & -0.000 & 0.005 & 0.43 & -0.06  & 0.004 & 0.007 & 0.050 & 0.09 & 0.13
\\ \cline{1-12}


\multirow{2}{*}{DES-Y1-5000, $w(\theta),\gamma_{\rm t}$} 
& $2\times 2$pt & 0.010 & -0.004 & 0.010 & 1.01 & -0.46 & -0.017 & 0.024 & 0.111 & -0.15 & 0.22
\\ \cline{2-12}
 & $3\times 2$pt & 0.003 & -0.001 & 0.006 & 0.62 & -0.19 & -0.007 & 0.006 & 0.050 & -0.14 & 0.12

\\\cline{1-12}
\end{tabular}
\egroup

\end{adjustbox}
\end{center}
\label{tab:S8values}
\end{table*}

In \sect\ref{sec:sensitivity} we showed the sensitivity of each of the two-point statistics, in terms of their signal-to-noise. Although this comparison gives us a measure of the systematic errors we can expect in the parameters estimated using each of these statistics, it does not tell the full story. Firstly, in \sect\ref{sec:sensitivity} we only considered the diagonal elements of the covariance matrix. For $C(\ell)$ that is a good approximation. However, both $\Psi$-statistics and correlation functions exhibit off-diagonal elements that can affect the final results. And secondly, to propagate the errors to the parameter space we also need to know the sensitivity of the statistics to these parameters. This is encoded in the Fisher matrix through their derivatives with respect to the parameters. Therefore, we use a Fisher matrix analysis to estimate the systematic errors on parameter estimation from each set of statistics. For our analysis we assume that the data comes from a scale-dependent galaxy bias model, while the theory used to analyse the data is based on a constant bias model. We can then write the displacement of the model parameters as,
\begin{equation}
\Delta\phi_p= \phi_p^{\rm fid}-\phi_p^{\rm est} = \sum_{qij} (\mathbfss{F}^{-1})_{p q}\ \frac{\partial\mu_i}{\partial\phi_q}\ (\mathbfss{C}^{-1})_{ij}\ (\mu_j - x_j)\;,
\label{eq:deltaphi}
\end{equation}
where $\phi_p^{\rm fid}$ is the true (fiducial) value of the parameter $\phi_p$ and $\phi_p^{\rm est}$ is its estimated value, $\mu_i$ is the theory prediction and $x_i$ is the data vector \citep{Taylor07}. The indices $p$ and $q$ indicate the model parameters, while $i$ and $j$ represent the elements of the data vector and its covariance matrix. We note that this formalism only provides the linear order terms which produce systematic errors.

For our analysis the data, $x_i$, comes from the scale-dependent galaxy bias models D18 and SH18. We then compute $\Delta\phi$ using \Eqt\eqref{eq:deltaphi}, but assuming a constant galaxy bias model for $\mu_i$.  As mentioned before we allow for five free cosmological parameters in our analysis: $\sigma_8$, $\Om$, $h$, $\Omega_{\rm b}$ and $n_{\rm s}$ as well as a number of galaxy bias parameters equal to the number of lens redshift bins. We marginalise over all parameters except for $\sigma_8$ and $\Om$, by removing their columns and rows from the parameter covariance matrix which is the inverse of the Fisher matrix.
We perform this analysis for both setups in \sect\ref{sec:surveys} with $\Psi$-statistics. For the KiDS-BOSS-like setup, we have seven free parameters and we compare the results to an analysis using the angular power spectra adopted by  \cite{vanuitert/etal:2018}, while for the DES-Y1-5000 setup we include ten free parameters and perform the comparison with the correlation functions adopted for the DES-Y1 3$\times$2pt analysis \citep{DES}. In order to achieve low sensitivity to the scale dependence of the galaxy bias, while retaining as much information as possible, we allow for freedom in both angular range and $n$-modes   included in the $\Psi$-statistics analysis. This choice is  justified, given that each survey adopted different levels of scale cuts to mitigate the impact of the scale-dependent galaxy bias.

We also show a cosmic shear prior based on the analysis of \cite{asgari/etal:2020} for both setups. The prior is produced by scaling the parameter covariance matrix obtained from the analysis of both KiDS-VIKING-450 and DES-Y1 data sets to the area of our two setups (see \tab\ref{tab:setups}). We combine the galaxy clustering and GGL results with the cosmic shear prior, but only over the two parameters of interest, $\sigma_8$ and $\Om$.

\fig\ref{fig:KiDS-BOSS-like_psi} shows the results for the KiDS-BOSS-like analysis. 
Here we chose $\theta\in[8',300']$ for the $\Psi$-statistics, with the value for $\theta_{\rm min}$ selected as a compromise between the size of the contours and the systematic bias in the parameter estimation. We use all $n$-modes between 1 and 20. 
For our $C(\ell)$ analysis we only include $\ell<1050$, similar to what is proposed for the analysis of the KiDS-1000 data\footnote{The scale-cuts for KiDS-1000 were later revised \citep[see][]{joachimi/etal:2020}}. We show the results for $2\times 2$pt, the combination of galaxy clustering and GGL as well as $3\times 2$pt which is the combination of $2\times 2$pt and the cosmic shear prior. The contours for the $\Psi$-statistics are larger than the ones from the Fourier space analysis. They are, however, much less affected by the non-linearity of the galaxy bias. 

The DES-Y1-5000 results are shown in \fig\ref{fig:DES-Y1-5000_psi}. The real space correlation functions have the same scale cuts as adopted by the DES-Y1 $3\times 2$pt analysis of \cite{DES} which, depending on the redshift bin, range from $\theta_{\rm min}=14'$ to $\theta_{\rm min}=64'$. Here we define the $\Psi$-statistics over the angular range of $[10',300']$ for all redshift bins using a $\theta_{\rm min}$ close to the minimum scale in the \cite{DES} real space analysis. To reduce the systematic biases, however, we exclude all $\Psi_n$ with $n>7$ from the analysis, since these modes are more sensitive to high $\ell$-scales (see \fig\ref{fig:WFilters}). The magenta and blue contours show the expected $\Delta\phi$ values for the $2\times 2$pt analysis, while the purple and green belong to the combination of the  $2\times 2$pt results and the cosmic shear prior shown in yellow. With the scale cuts used here we see that the constraining power of the $\Psi$-statistics is higher than the conservatively cut real space correlation functions.

\subsection{Quantifying parameter bias in $S_8$}
The results of a  $3\times2$pt analysis is usually quoted in terms of a combination of $\sigma_8$ and $\Om$ for example $S_8=\sigma_8(\Om/0.3)^{\alpha}$. This combination is defined such that it captures the degeneracy direction of the data, resulting is smaller errors compared to either $\sigma_8$ or $\Om$. The value of $\alpha$ depends on the data that is being analysed, although most recent analyses fix it to 0.5. Given that we use a Fisher analysis, we define two new parameters based on linear combination of $\sigma_8$ and $\Om$. These parameters are $\S8$  and   $\Sp$ which are defined along the minor and major axis of the error ellipses, respectively. We can relate $\S8$ to $S_8$ around $\Omega_{\rm}=0.3$ using a Taylor expansion,
\begin{equation}
\label{eq:taylorexp}
\sigma_8\approx (1+\alpha)S_8 - \alpha S_8 \Om/0.3\;,
\end{equation}
and solve for $\alpha$. We find $\alpha$ values ranging from $0.6$ to $0.9$. \tab\ref{tab:S8values} shows values for $\DS8$  and   $\DSp$ as well as their standard deviation, $\sigS8$ and $\sigSp$, for the setups used in \figs\ref{fig:KiDS-BOSS-like_psi} and \ref{fig:DES-Y1-5000_psi}.

For the KiDS-BOSS-like setup we see that the $C(\ell)$ analysis results in smaller errors but larger biases. Although the biases on $\S8$ are not very large ($<0.5\sigma$), the biases on 
$\Sp$ are significant ($2.2\sigma$ to $3.8\sigma$). 
This shows that a KiDS-BOSS-like $3\times 2$pt analysis will need to go beyond scale-independent galaxy bias modelling for angular power spectra defined over $150<\ell<1050$.
The values of $\DS8$ for the $\Psi$-statistics are either equal to, or smaller than, the Fourier analysis.  The errors, however, are roughly twice as large as its $C(\ell)$ counterpart for the angular range adopted.   This results in significantly smaller systematic biases on the inferred parameters. In particular, for $\S8$ we have systematic errors of $\sim 0.1\sigma$.

The DES-Y1-5000 results in \tab\ref{tab:S8values} show that with $\Psi$-statistics we can reduce $\DS8$ by half, reduce the overall errors, and also find smaller systematic biases when compared to the real space analysis which can show up to $1\sigma$ systematic errors on $\S8$. 
Turning to $\Sp$ we see that the systematic biases are not significant, as there is no
strong degeneracy breaking from the clustering signal. This can be seen in \fig\ref{fig:DES-Y1-5000_psi} which shows elongated ellipses with very different sizes for their minor and major axes. Comparing that figure with \fig\ref{fig:KiDS-BOSS-like_psi} we see that in the $C(\ell)$ analysis of the KiDS-BOSS-like setup there is significant degeneracy breaking from the clustering signal of BOSS and therefore the contours are less elliptical.

Finally, comparing the $2\times2$pt and $3\times2$pt results in  \tab\ref{tab:S8values},  we see that combining the data with cosmic shear always improves results by decreasing the biases.   This is expected, since cosmic shear is insensitive to galaxy bias, highlighting a key benefit of the joint cosmological analysis of large-scale structures.

%% file: Conclusions.tex
\label{sec:Conclusions}
In this paper we introduced the $\Psi$-statistics, which are two-point statistics designed for combining galaxy-galaxy lensing and galaxy clustering analysis. 
We compared them with the traditionally used statistics: real space correlation functions, $w(\theta)$ and $\gamma_{\rm t}(\theta)$, as well as Fourier space angular power spectra, $C(\ell)$. The $\Psi$-statistics are inspired by COSEBIs \citep{SEK10}. They are defined as integrals over the real space correlation functions with filters specified on a finite angular range. We can measure an unbiased estimate of $\Psi$-statistics from the data, by choosing an angular range where measurements of correlation functions are available. This can be an issue for $C(\ell)$ as they are usually\footnote{There are other methods to calculate power spectra, for example using a quadratic estimator \citep{kohlinger/etal:2017}, which are sensitive to the accuracy of noise modelling.} either calculated by integrating over correlation functions on an infinite range of $\theta$-scales \citep{vanuitert/etal:2018} or by Fourier transforming the field which produces pseudo-Cls that can be biased by masking effects \citep{asgari/etal:2018}.

 $\Psi$-statistics are formed of discrete and well-defined modes, unlike the traditional statistics which need to be binned,  complicating the analysis and the covariance estimation \citep{Troxel/etal:2018b, asgari/etal:2019}. The $\Psi$-statistics,  much like COSEBIs, limit the $\ell$-dependence of the measurements to large $\ell$-modes. For example, considering the angular range of $[0.5', 300']$ we find that the range of support of $\Psi$-statistics is between $\ell_{\rm min}\approx 10$ and  $\ell_{\rm max}$ of a few hundred (see \fig\ref{fig:integrands}). While with $w(\theta)$ the range of support is much larger, from $\ell$ of a few to more than 10,000. The correlation function, $\gamma_{\rm t}$, on the other hand, has a more compact weight function, but generally probes larger $\ell$-scales. 

When galaxies are used as tracers for the matter distribution, we need to include galaxy bias modelling in the analysis. The galaxy bias is believed to be constant on very large scales for each population, but have a scale-dependent form for smaller scales. The scale at which this scale-dependence becomes important and its shape, depends on the galaxy population. In this paper we investigated the effect of two scale-dependent galaxy bias models on a cosmological analysis of the large scale structures based on models adapted from \citet[D18]{Andrej_bias_model}  and \citet[SH18]{Simon_bias_model}. We combined galaxy clustering, galaxy-galaxy lensing and cosmic shear, to form a $3\times 2$pt analysis of the large scale structures. Most analyses of this kind assume a constant bias model with a free amplitude and apply scale cuts to limit the contaminations from the scale-dependent biases \citep{DES,vanuitert/etal:2018,Joudaki_KiDS_2dFLenS}. Here we chose two survey setups similar to the combination of KiDS-1000 and BOSS, the KiDS-BOSS-like setup and DES-Y1-5000 setup, based on the DES-Y1 survey  but scaled to match the area of the final DES data, as described in \sect\ref{sec:surveys}. Using these setups we test this assumption.

We compared the sensitivity of $\Psi$-statistics to scale-dependent galaxy bias with $C(\ell)$ and the combinations of $w(\theta)$ and $\gamma_{\rm t}(\theta)$. We first considered the KiDS-BOSS-like setup, computed the difference between signal-to-noise ratios for each scale-dependent model and the constant bias case and then compared these values between the different sets of statistics. We found that over the angular range of $[0.5',300']$, $\Psi$-statistics are far less sensitive to non-linear bias (up to $0.9\sigma$), compared to the real space correlation functions (up to $60\sigma$), but they contain essentially the same cosmological information  as in correlation functions (see \fig\ref{fig:Figure_Of_Merit}). For the $C(\ell)$ analysis we chose the range $150\leq\ell\leq2000$, where we found that the signal-to-noise ratio for the angular power spectra are affected significantly (up to $18\sigma$), although not as much as $w(\theta)$ and $\gamma_{\rm t}(\theta)$. The clustering signal is generally more affected, as it scales with the square of the bias functions, $b(k)$, which in most models has a more significant non-linear dependence compared to the bias function, $r(k)$. The galaxy-galaxy lensing signal probes the combination of these two bias functions,  $r(k)b(k)$. 

We used a Fisher analysis to propagate the systematic errors, introduced by ignoring the scale dependence of the galaxy bias, to cosmological parameters. For this analysis we assume that the data comes from one of the two scale-dependent bias models, while the analysis is performed with a constant bias model. We allowed for five cosmological parameters: $\sigma_8$, $\Om$, $h$, $\Omega_{\rm b}$ and $n_{\rm s}$ as well as a number of effective galaxy bias parameter equal to the number of lens bins to vary. We showed marginalised results for $\sigma_8$ and $\Om$. Our KiDS-BOSS-like setup is the closest to the KiDS-1000 combined probe analysis which will be performed with band power spectra. Therefore, we compared $\Psi$-statistics defined on $[8',300']$ with $C(\ell)$ defined on $150\leq\ell\leq1050$, a range that is close to what will be used in the KiDS-1000 $\times$ BOSS analysis. 
The DES-Y1 combined probe analysis employed correlation functions, $w(\theta)$ and $\gamma_{\rm t}(\theta)$, with conservative scale cuts to reduce the effect of non-linear galaxy bias. With our DES-Y1-5000 setup we predict the systematic errors that are expected for the full data analysis of DES but only up to the depth of the year 1 result. We used the same conservative scale cuts for the correlation functions and compared the results with $\Psi$-statistics defined for $[10',300']$ with $n<8$.

We quantified the systematic errors for $\S8$ and $\Sp$ two parameters defined along the minor and major axes of the error ellipses for $\sigma_8$ and $\Om$, respectively.  $\S8$ is defined to be similar to $S_8=\sigma_8(\Om/0.3)^{\alpha}$, but also more relevant for a Fisher analysis. We summarise the results in \tab\ref{tab:S8values}. For the  KiDS-BOSS-like analysis we see that the systematic errors, $\DS8$ and $\DSp$, are smaller in the case of $\Psi$-statistics. But the parameter errors, $\sigS8$ and $\sigSp$, are larger and therefore the relative value of the systematic error to parameter error is smaller for $\Psi$-statistics, which means that they are less affected by the scale-dependence of galaxy bias. Interestingly, we see that although the systematic biases on $\Sp$ is large for the Fourier analysis, the value of $\S8$ is less significantly biased.

For the DES-Y1-5000 analysis we can decrease systematic errors from the $\Psi$-statistics on $\DS8$ by factors of 2 to 4, dependent on the galaxy bias model adopted. With the angular range that we chose for the $\Psi$-statistics we get slightly tighter constraints compared to the real space analysis. Nevertheless, the relative sizes of the systematic error to the parameter error, $\DS8/\sigS8$, is smaller for the the $\Psi$-statistics, making them less affected by the scale-dependence of galaxies bias. In contrast to the KiDS-BOSS-like analysis we see a larger bias on the value of $\S8$, while $\Sp$ is practically unaffected. This is due to degeneracy braking along $\Om$ from the clustering signal in the BOSS data, which is not present in the DES-Y1-5000 with the conservative scale-cuts. However, with the DES-Y1-5000 we have more GGL area which results in much tighter  $\S8$ constraints.
Given that $\S8$ is closer to the quantity of interest, $S_8$, it is important to reduce its dependency on the modelling of  galaxy bias. The final DES data will be deeper than the setup we have considered here and the scale-dependence of galaxy bias will likely become even more important for that analysis. With optimisation we anticipate being able to further decrease the $\Psi$-statistic sensitivity to scale-dependent galaxy bias.  For example, with COSEBIs there are two sets of families with linear and logarithmic filter functions, where the logarithmic functions are able to compress the data into fewer modes. In future  work we will apply logarithmic filters to $\Psi$-statistics to make them more efficient, which we expect will also make them even less sensitive to galaxy bias. 

We conclude by reminding the reader that the galaxy bias models that we have used are reasonable toy models that become unreliable at high $k>10\,h\,{\rm Mpc}^{-1}$. We do not have, and probably will not be able to obtain, the data or simulations to accurately model these scales.  We therefore argue that the community should move away from using statistics that are sensitive to high $k$ values, especially those that significantly mix different $k$ values. The $\Psi$-statistics provide a promising new alternative with reduced sensitivity to the scale-dependence of galaxy bias.

%% file: covariance.tex
\label{app:covariance}

The covariance of $\Psi_n$ and $\Psi_m$ for pairs of redshift bins, $ij$ and $kl$ can be written in terms of the covariance of projected power spectra, $C_{{\rm wxyz}}(\ell,\ell')$,
\begin{align}
{\rm Cov}(\Psi_{{\rm wx},n}^{ij}, \Psi_{{\rm yz}, m}^{kl})= C^{ijkl}_{{\rm wxyz},nm}= 
\left\langle \Psi_{{\rm wx},n}^{ij} \Psi_{{\rm yz}, m}^{kl} \right\rangle - \left\langle \Psi_{{\rm wx},n}^{ij} \right\rangle  \left\langle \Psi_{{\rm yz}, m}^{kl} \right\rangle =
\int_0^\infty \frac{\d \ell\;\ell}{2\pi} \int_0^\infty \frac{\d \ell'\;\ell'}{2\pi} W_n(\ell) W_m(\ell') C^{ijkl}_{{\rm wxyz}}(\ell,\ell')\;,
\end{align}
where w, x, y and z stand for either g (galaxy) or m (matter). Since our data vector comprises galaxy clustering and GGL correlations, we are only interested in four combinations of these indices, gggg for clustering, gmgm for galaxy-galaxy lensing and gggm or gmgg for their cross-covariance.  Assuming that only Gaussian terms contribute to the covariance we can write these terms as,
\begin{align}
\label{eq:cov_all}
C^{ijkl}_{{\rm gggg},nm} &= \frac{1}{2\pi A}\int_0^\infty \d \ell\; \ell\; W_n(\ell) W_m(\ell) 
\left[ \bar{C}^{ik}_{\rm gg}(\ell)\bar{C}^{jl}_{\rm gg}(\ell) +  \bar{C}^{il}_{\rm gg}(\ell)\bar{C}^{jk}_{\rm gg}(\ell) \right]\;, \\ \nonumber
C^{ijkl}_{{\rm gmgm},nm} &= \frac{1}{2\pi A}\int_0^\infty \d \ell\; \ell\; W_n(\ell) W_m(\ell) 
\left[ \bar{C}^{ik}_{\rm gg}(\ell)\bar{C}^{jl}_{\rm mm}(\ell) +  C^{il}_{\rm gm}(\ell)C^{jk}_{\rm mg}(\ell) \right]\;, \\ \nonumber
C^{ijkl}_{{\rm gggm},nm} &= \frac{1}{2\pi A}\int_0^\infty \d \ell\; \ell\; W_n(\ell) W_m(\ell) 
\left[ \bar{C}^{ik}_{\rm gg}(\ell) C^{jl}_{\rm gm}(\ell) +  C^{il}_{\rm gm}(\ell)\bar{C}^{jk}_{\rm gg}(\ell) \right]\;, 
\end{align}
where
\begin{equation}
\bar{C}^{ij}_{\rm gg}(\ell)= C^{ij}_{\rm gg}(\ell)+\frac{1}{\bar{n}_i}\delta_{ij}, ~~~~~~
\bar{C}^{ij}_{\rm mm}(\ell)= C^{ij}_{\rm mm}(\ell)
+\frac{\sigma^2_{\epsilon,i}}{2\bar{n}_i} \delta_{ij}\;.
\end{equation}
Here $C^{ij}_{\rm gg}(\ell)$, $C^{ij}_{\rm gm}(\ell)$ and $C^{ij}_{\rm mm}(\ell)$ are the projected galaxy, galaxy-matter and matter power spectra for redshift bin pair $i$ and $j$, respectively. The number density of galaxies is given by $\bar{n}_i$ for redshift bin $i$, $\delta_{ij}$ is the Kronecker delta and $\sigma_\epsilon$ denotes the intrinsic dispersion of galaxy ellipticities.  The gmgg terms are simply the transpose of the gggm terms. We note that in \Eqt\ref{eq:cov_all} the cross power spectrum terms, $C^{ij}_{\rm gm}(\ell)$ and  $C^{ij}_{\rm mg}(\ell)$, do not include noise.

\begin{figure}
\begin{center}
     \begin{tabular}{c}
  \includegraphics[width=0.5\linewidth]{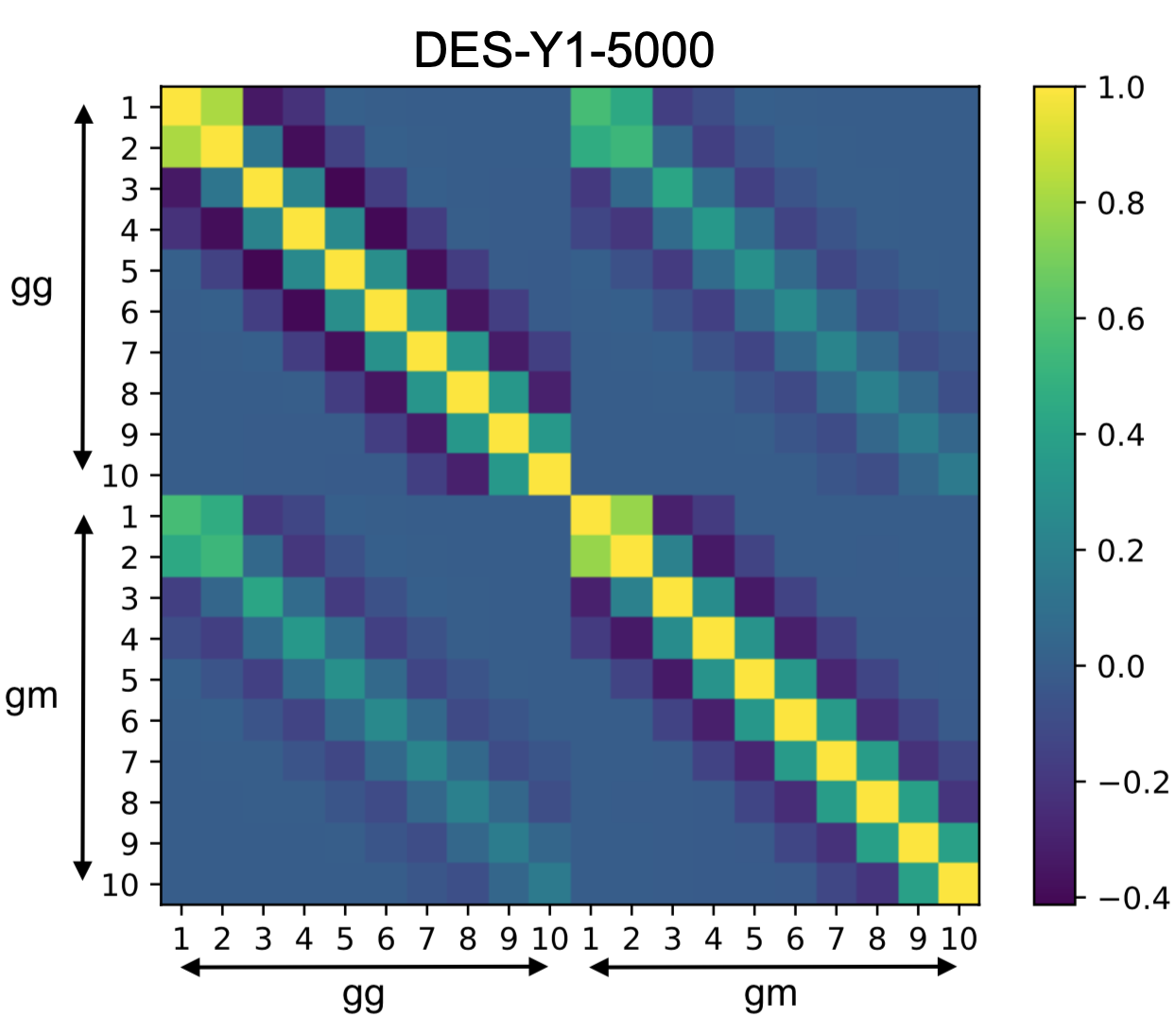}
  \end{tabular}
  \end{center}
\caption{{\small Correlation matrix for $\Psi$-statistics. Results are shown for the clustering signal for the auto-correlation of the first redshift bin (upper left), its GGL signal with the fourth source redshift bin (lower right) and their cross-covariance (lower left and upper right) for the DES-Y1-5000 setup. We only show values for $1\leq n\leq 10$.} }
\label{fig:covariance}
\end{figure}

\fig\ref{fig:covariance} shows part of the $\Psi$-statistics correlation matrix for the DES-Y1-5000 setup (see \sect\ref{sec:surveys}). Here we only show correlations with lens bin 1 and source bin 4 with 10 modes defined over the angular range $[0.5', 250']$. The top left block shows the clustering correlation matrix corresponding to the gggg term in \Eqt\ref{eq:cov_all}, while the lower right block shows the gmgm terms which is the correlation matrix for GGL. The two off-diagonal blocks show the cross-correlation matrices between clustering and GGL, which are the gggm and gmgg terms.

%% file: appendix_fig.tex

%

\label{app:fig}

\begin{figure*}
	\includegraphics[width=\textwidth]{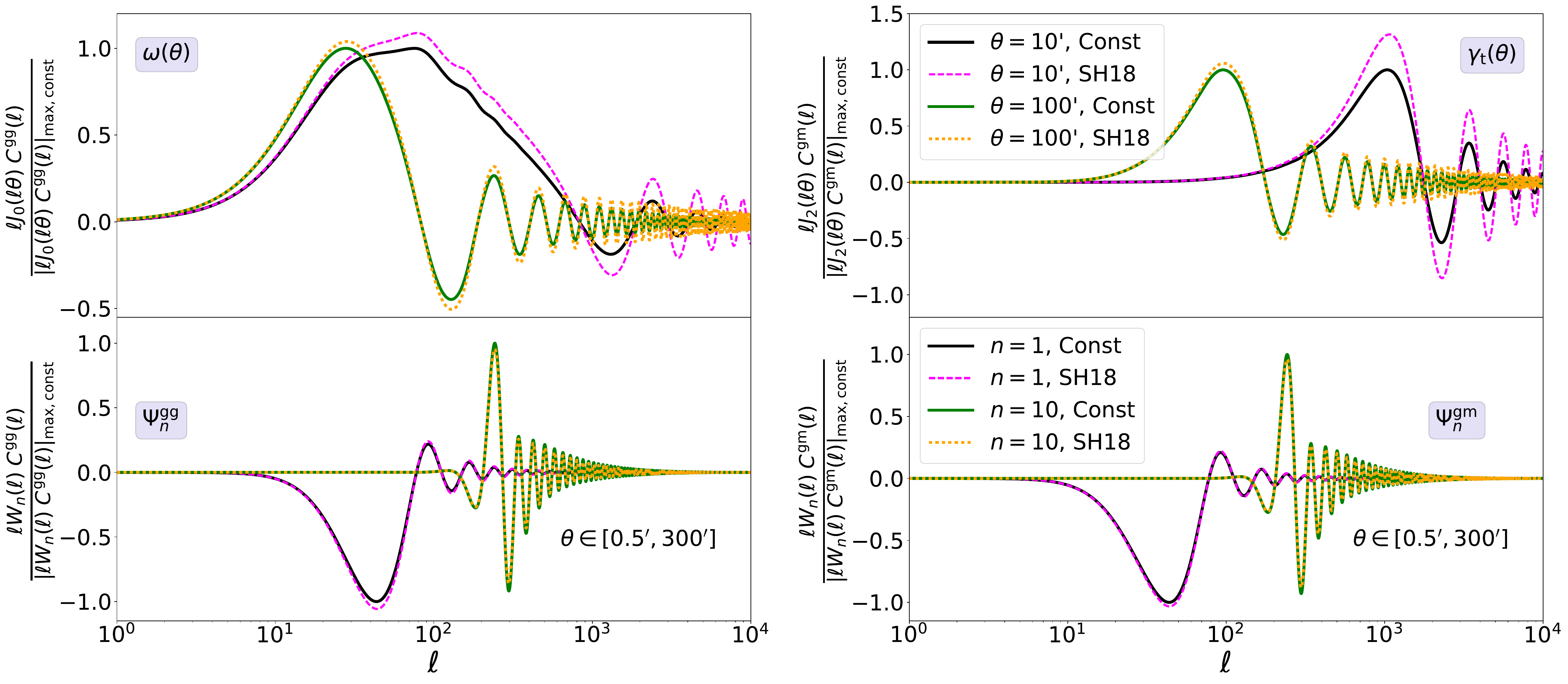}
    \caption{\small{Integrands of correlation functions and $\Psi$-statistics using constant and the SH18 scale-dependent galaxy bias model, similar to \fig\ref{fig:integrands}. Here we show $\omega(\theta)$ (left) and $\gamma_{\rm t}(\theta)$ (right) for $\theta=10'$ and $\theta=100'$, while the $\Psi^{\gg}_{n}$ (left) and  $\Psi^{\gm}_{n}$ (right) integrands are shown for both  $n$-mode=1 and $n$-mode=10 defined over an angular range of $[0.5',300']$. All integrands are normalised by their absolute maximum value for the constant bias case.}}
    \label{fig:integrands_app}
\end{figure*}

 In \fig\ref{fig:integrands} we presented the integrands that convert power spectra to real space correlations and $\Psi$-statistics. That figure shows results for both SH18 and D18 models, but given a single angular scale and $n$-mode. With \fig\ref{fig:integrands_app} we contrast the integrands for two angular scales, $\theta=10'$ and $\theta=100'$, as well as two modes, $n=1$ and $n=10$. We see that even at larger $\theta$ the real space correlation functions are still sensitive to larger $\ell$-scale where the scale-dependence of galaxy bias becomes important.